\documentclass[twocolumn, trackchanges, twocolappendix]{aastex631}

\usepackage{CJK}
\usepackage{ulem}
\usepackage{mathptmx}
\usepackage{hyperref}
\hypersetup{bookmarksnumbered=true}

\newcommand\hi{{\rm H}{\textsc i}}
\newcommand{\dsfr}{$\Delta {\log{\rm SFR}}$}
\newcommand{\dproj}{$d_{\rm proj}$}
\newcommand{\dv}{$\Delta v_{\rm rad}$}
\newcommand{\dvn}{\dv$/\sigma_{v,\rm pair}$}
\newcommand{\dprojn}{\dproj$/R_{\rm 200,pair}$}
\newcommand{\rgas}{$r_{\rm gas}$}

\shorttitle{SFR of WALLABY galaxy pairs}
\shortauthors{Huang et al.}

\graphicspath{{./}{figures/}}

\begin{document}
\begin{CJK*}{UTF8}{gbsn}

\title{WALLABY Pilot Survey: Star Formation Enhancement and Suppression in Gas-rich Galaxy Pairs}

\correspondingauthor{Jing Wang}
\email{jwang\_astro@pku.edu.cn}

\author[0000-0003-2863-9837]{Qifeng Huang (黄齐丰)}
\affiliation{Kavli Institute for Astronomy and Astrophysics, Peking University, Beijing 100871, People's Republic of China}
\affiliation{Department of Astronomy, School of Physics, Peking University, Beijing 100871, People's Republic of China}

\author[0000-0002-6593-8820]{Jing Wang (王菁)}
\affiliation{Kavli Institute for Astronomy and Astrophysics, Peking University, Beijing 100871, People's Republic of China}

\author[0000-0002-4250-2709]{Xuchen Lin (林旭辰)}
\affiliation{Kavli Institute for Astronomy and Astrophysics, Peking University, Beijing 100871, People's Republic of China}
\affiliation{Department of Astronomy, School of Physics, Peking University, Beijing 100871, People's Republic of China}

\author[0000-0002-8379-0604]{Se-Heon Oh}
\affiliation{Department of Physics and Astronomy, Sejong University, 209 Neungdong-ro, Gwangjin-gu, Seoul, Republic of Korea}

\author[0000-0002-5016-6901]{Xinkai Chen (陈新凯)}
\affiliation{Kavli Institute for Astronomy and Astrophysics, Peking University, Beijing 100871, People's Republic of China}

% WALLABY builder list
\author[0000-0002-7625-562X]{B. Catinella}
\affiliation{International Centre for Radio Astronomy Research (ICRAR), The University of Western Australia, 35 Stirling Highway, Crawley WA 6009, Australia}
\affiliation{ARC Centre of Excellence for All Sky Astrophysics in 3 Dimensions (ASTRO 3D), Australia}

\author[0000-0003-3523-7633]{N. Deg}
\affiliation{Department of Physics, Engineering Physics and Astronomy, Queen's University, Kingston, ON K7L~3N6, Canada}

\author[0000-0002-9214-8613]{H. D\'enes}
\affiliation{School of Physical Sciences and Nanotechnology, Yachay Tech University, Hacienda San Jos\'e S/N, 100119, Urcuqu\'i, Ecuador}

\author[0000-0002-0196-5248]{B.~-Q. For}
\affiliation{International Centre for Radio Astronomy Research (ICRAR), The University of Western Australia, 35 Stirling Highway, Crawley WA 6009, Australia}
\affiliation{ARC Centre of Excellence for All Sky Astrophysics in 3 Dimensions (ASTRO 3D), Australia}

\author[0000-0003-4351-993X]{B. S. Koribalski}
\affiliation{Australia Telescope National Facility, CSIRO, Space and Astronomy, P.O. Box 76, Epping, NSW 1710, Australia}
\affiliation{School of Science, Western Sydney University, Locked Bag 1797, Penrith, NSW 2751, Australia}

\author[0000-0003-4844-8659]{K. Lee-Waddell}
\affiliation{International Centre for Radio Astronomy Research (ICRAR), The University of Western Australia, 35 Stirling Highway, Crawley WA 6009, Australia}
\affiliation{CSIRO Space and Astronomy, PO Box 1130, Bentley WA 6102, Australia}
\affiliation{International Centre for Radio Astronomy Research (ICRAR), Curtin University, Bentley, WA 6102, Australia}

\author[0000-0001-8496-4306]{J. Rhee}
\affiliation{International Centre for Radio Astronomy Research (ICRAR), The University of Western Australia, 35 Stirling Highway, Crawley WA 6009, Australia}

\author{A. X. Shen}
\affiliation{CSIRO Space and Astronomy, PO Box 1130, Bentley WA 6102, Australia}

\author[0000-0003-2015-777X]{Li Shao (邵立)}
\affiliation{National Astronomical Observatories, Chinese Academy of Sciences, Beijing 100101, China}

\author[0000-0002-0956-7949]{K. Spekkens}
\affiliation{Department of Physics, Engineering Physics and Astronomy, Queen's University, Kingston, ON K7L~3N6, Canada}

\author[0000-0002-8057-0294]{L. Staveley-Smith}
\affiliation{International Centre for Radio Astronomy Research (ICRAR), The University of Western Australia, 35 Stirling Highway, Crawley WA 6009, Australia}
\affiliation{ARC Centre of Excellence for All Sky Astrophysics in 3 Dimensions (ASTRO 3D), Australia}

\author[0000-0002-5300-2486]{T. Westmeier}
\affiliation{International Centre for Radio Astronomy Research (ICRAR), The University of Western Australia, 35 Stirling Highway, Crawley WA 6009, Australia}

\author[0000-0003-4264-3509]{O. I. Wong}
\affiliation{CSIRO Space and Astronomy, PO Box 1130, Bentley WA 6102, Australia}
\affiliation{International Centre for Radio Astronomy Research (ICRAR), The University of Western Australia, 35 Stirling Highway, Crawley WA 6009, Australia}
\affiliation{ARC Centre of Excellence for All Sky Astrophysics in 3 Dimensions (ASTRO 3D), Australia}

\author[0000-0002-1128-6089]{A. Bosma}
\affiliation{Aix Marseille Universit\'e, CNRS, CNES, LAM, F-13388 Marseille Cedex 13, France}

\begin{abstract}
Galaxy interactions can significantly affect the star formation in galaxies, but it remains a challenge to achieve a consensus on the star formation rate (SFR) enhancement in galaxy pairs. Here, we investigate the SFR enhancement of gas-rich galaxy pairs detected by the Widefield ASKAP $L$-band Legacy All-sky Blind surveY (WALLABY). We construct a sample of 278 paired galaxies spanning a stellar mass ($M_\ast$) range from $10^{7.6}$ to $10^{11.2}M_\odot$. We obtain individual masses of atomic hydrogen (\hi) for these paired galaxies, using a novel deblending algorithm for \hi\ data cubes. Quantifying the interaction stages and strengths with parameters motivated by first principles, we find that at fixed stellar and \hi\ mass, the alteration in SFR of galaxy pairs starts when their dark matter halos encounter. For galaxies with stellar mass lower than $10^9M_\odot$, their SFRs show tentative suppression of $1.4\sigma$ after the halo encounter, and then become enhanced when their \hi\ disks overlap, regardless of mass ratios. In contrast, the SFRs of galaxies with $M_\ast > 10^9M_\odot$ increase monotonically toward smaller projected distances and radial velocity offsets. When a close companion is present, a pronounced SFR enhancement is found for the most \hi-poor high-mass galaxies in our sample. Collecting the observational evidence, we provide a coherent picture of the evolution of galaxy pairs, and discuss how the tidal effects and hydrodynamic processes shape the SFR enhancement. Our results provide a coherent picture of gas-rich galaxy interactions and impose constraints on the underlying physical processes.
\end{abstract}


\section{Introduction} \label{sec:intro}

In a hierarchical universe, galaxies do not live alone, but interact with their neighbors. The interaction between galaxies has a significant impact on their evolution. For example, the structure and morphology of galaxies can be easily altered by close encounters, leaving asymmetric features commonly found in observations \citep{Arp1966, Toomre1972, Huang2022} and transforming disks into spheroidals \citep{Barnes1992}. Gas content and flows within interacting galaxies can also be disrupted, which leads to intricate consequences such as the redistribution of metals \citep{Kewley2010, Hani2018, Perez-Diaz2024}, the transformation of gas phases \citep{Pan2018, Moreno2019, Wang2023a}, the altering of the star formation states \citep{Ellison2008, Patton2013, Gao2023, Kado-Fong2024, Bottrell2023}, and the triggering of active galactic nuclei (AGNs) activities \citep{Silverman2011, Satyapal2014, Ellison2019}.

The connection between galaxy interactions and their star formation rates (SFRs) has long been studied. Violent mergers are thought to be the main channel of starbursts in the local Universe, as supported by the morphological disturbances seen around the majority of these systems \citep{Conselice2003, Pawlik2016}. Besides, mergers are likely responsible for the quenching of galaxies through depleting the cold gas or preventing the existing gas from forming stars \citep{Martig2009,Ellison2022,Petersson2022,Quai2023,Kado-Fong2024}. More subtle effects on the SFR would happen during the interaction phase well before mergers. When selecting galaxy pairs in observations based on the proximity of projected distances and radial velocities, previous studies found that the SFR of interacting galaxy pairs increases as their distances decrease, compared to isolated samples (\citealt{Ellison2008,Moon2019,Pan2019,Garduno2021}, but see \citealt{Li2023a} and \citealt{Garay-Solis2023} for opposite results). Numerical simulations of galaxy mergers successfully reproduce this phenomenon, which typically trace the SFR enhancement as a function of time for individual merging systems \citep[e.g.,][]{Moreno2019, Sparre2022, Renaud2022}. Cosmological simulations showed with better statistics that the SFR enhancement becomes significant even at separations over 200 kpc \citep{Patton2020}, indicating a long-lasting impact on the SFR during the galaxy pair periods.

Despite the general trend of interacting galaxies, the detailed behavior of the SFR enhancement depend intricately on the properties of the galaxies involved as well as on the environment they inhabit. As the fuel of star formation, cold gas content of the progenitor galaxies can substantially influence how the SFR varies during the interactions. For instance, star formation in gas-rich galaxies is more enhanced when the neighbors are also gas-rich and star-forming \citep{Xu2010, Zuo2018, Moon2019, Brown2023a}. Using hydrodynamic simulations, \citet{Scudder2015} found that both the time and magnitude of the SFR enhancement are sensitive to the gas fractions of the progenitors. Observations have also showed that the presence of a second companion or massive galaxies nearby can suppress the SFR enhancement \citep{Stierwalt2015,Bustamante2020}. In more extreme situations, satellite galaxies falling into massive halos could be quenched by the environment \citep{Cortese2021, Engler2023}. In addition, the merger mass ratio also matters. Gas-rich major mergers are found to induce the most prominent starbursts \citep{Woods2006, Pan2018}, and minor mergers that are preponderant in number may have the largest contribution to the interaction-induced star formation budget \citep{Bottrell2023}. In terms of physical mechanisms, although mergers and tidal interactions are driven by gravity, hydrodynamic effects can also affect the SFRs, especially for the gas-rich systems we consider. These effects can also lead to both SFR elevation and suppression \citep[e.g.,][]{Moon2019, Spilker2022, Kado-Fong2024}, depending on the specific situations. All these complexities urge us to put together the parameter dependencies coherently in a physically motivated picture, decompose the different mechanisms possibly affecting the SFR enhancement, and clarify the role of gas during the process.

Previous observational studies usually use the projected distances and the mass ratios to quantify the interaction strength. Although highly intuitive and easy to obtain in observations, these parameters provide limited insight into the physical mechanisms in action. Furthermore, the proximity of galaxy pairs in the sky poses challenges for separating their gas content with blind \hi\ surveys using single-dish telescopes, such as the Arecibo Legacy Fast ALFA survey \citep[ALFALFA;][]{Giovanelli2005} and the FAST All Sky \hi\ survey \citep[FASHI;][]{Zhang2024}. As a result, the sample size of galaxy pairs with unblended \hi\ detection is consistently smaller compared to optically selected samples \citep[e.g.,][]{Scudder2015}, impeding the study of atomic gas in interacting galaxies. 

The ongoing Widefield ASKAP $L$-band Legacy All-sky Blind surveY \citep[WALLABY;][]{Koribalski2020} is observing the southern sky to map the \hi\ of galaxies in the local Universe with a spatial resolution of $30''$. It provides a much more advantageous dataset than before to construct a large sample of interacting galaxies with spatially resolved \hi\ properties. In this study, we utilize the data from WALLABY to define our sample of gas-rich galaxy pairs in the local Universe. We also define parameters directly linked to different physical mechanisms (e.g., tidal perturbation and collision between the gas disks) and investigate how they contribute to the SFR enhancement. As WALLABY is covering more fields \citep[e.g.,][]{Wong2021,Reynolds2022,For2023} and detecting an increasing number of interacting systems, the method presented here could be applied to much larger samples to obtain better statistics and reveal more subtle effects of galaxy interactions. 

This paper is arranged as follows. Section \ref{sec:data} details the data employed in our study and the method to build the galaxy samples. Section \ref{sec:deblend} introduces the technique we developed to deblend gas-rich galaxy pairs in \hi\ data cubes. The main results are described in Section \ref{sec:results}. In Section \ref{sec:discuss}, we provide a thorough discussion on the evolution of galaxy pairs based on the results, depicting the evolutionary tracks of the gas-rich galaxy pairs. Finally, we conclude the paper in Section \ref{sec:conclusion}. Throughout this paper, we adopt a standard Lambda cold dark matter ($\Lambda$CDM) cosmology with $H_0=70~\rm{km~s^{-1}~Mpc^{-1}}$, $\Omega_{\rm m,0}=0.30$, and $\Omega_{\Lambda,0}=0.70$. We assume a \citet{Chabrier2003} stellar initial mass function. All the magnitudes are given in the AB system. 

% ------------------------
\section{Data and sample} \label{sec:data}

\subsection{\hi-rich galaxies in WALLABY} \label{subsec:wallaby}

WALLABY is an ongoing all-sky \hi\ blind survey aiming to cover three-quarters of the sky out to a redshift of $z\approx 0.1$ \citep{Koribalski2020, Westmeier2022, Murugeshan2024}. The \hi\ data from WALLABY reaches a spatial resolution of $30''$, with a channel width of about 4 $\rm km~s^{-1}$. The detection limit of the data is $\lesssim 2~\rm mJy~beam^{-1}$ except at the edge of the fields, corresponding to a $3\sigma$ \hi\ column density limit of $6.5\times10^{19}\rm~cm^{-2}$ over 5 channels. \hi\ source detection is carried out with \texttt{SoFiA} \citep{Serra2015, Westmeier2021}, which provides a wealth of information including radial velocities, coordinates, \hi\ total fluxes, and \hi\ line widths. 

We begin our analysis with the galaxies with \hi-detection in the fields observed during the pilot phases of WALLABY, including the Hydra, NGC 4636, NGC 4808, and NGC 5044 fields \citep{Westmeier2022, Murugeshan2024}, with a total sky coverage of $\sim 270~{\rm deg}^2$ (see also Table \ref{tab:group}). We visually inspect all the 1976 sources detected by \texttt{SoFiA} taking the advantage of optical images from the DESI Legacy Imaging Surveys \citep{Dey2019}. We exclude 136 sources with no definite optical counterparts,  corresponding to only half of a galaxy, or contaminated by continuum artifacts. We also pick out galaxy pairs and triplets treated as single \hi\ sources by \texttt{SoFiA}. For these blended systems, we develop a deblending algorithm to split the \hi\ total flux into each galaxy (described in Section \ref{sec:deblend}). Two blended galaxy pairs, WALLABY J103442-283406 and WALLABY J123424+062511, have been excluded from our sample because attributing the \hi\ in extended gas tails and clouds to either galaxy proved challenging (Appendix \ref{app:hi_tail}). These two systems are studied in separate projects (\citealt{OBeirne2024}; Staveley-Smith et al. in prep). Finally, a total of 1922 \hi-detected galaxies are selected from the WALLABY data.

\subsection{Selection of galaxy pairs} \label{subsec:pair}

We select galaxy pairs from all the 1922 gas-rich galaxies obtained in Section \ref{subsec:wallaby}. Details of the selection criteria are described as follows.

\subsubsection{Selection based on projected distance and radial velocity offset} \label{subsubsec:d&v}
In observations, galaxy pairs are often selected based on their proximity in projected distances (\dproj) and radial velocities ($v_{\rm rad}$). Utilizing the IllustrisTNG simulations, \cite{Patton2020} showed that the SFR enhancement of galaxy pairs becomes statistically significant at separations smaller than $\sim$ 250 kpc, comparable to the virial radius of the dark matter halo of a Milky Way-like galaxy. In observational studies, smaller separations are required to detect significant effects of galaxy interactions \citep[e.g.,][]{Patton2013, Bustamante2020}. Based on these results, we select galaxy pairs with $d_{\rm proj} < 250$ kpc to focus on galaxies likely experiencing SFR enhancement.

We also require the radial velocity offset between the two galaxies to satisfy \dv\ $<500~\rm km~s^{-1}$. This criterion is more relaxed than those commonly used in the literature \citep[e.g.,][]{Ellison2008, Kim2023}, enabling us to construct a more complete galaxy pair sample and incorporate galaxy flybys in addition to mergers \citep{Moreno2013, Cerdosino2024}. The flybys will not contaminate our sample when considering the scientific goals, since the effects of ongoing interactions are agnostic of whether the systems will merge eventually. Instead, they help to expand the dynamic range of interaction strength between galaxy pairs.

There are 631 galaxies left after applying the cuts on \dproj\ and \dv, of which 368 are in pairs and 263 have multiple companions. We will further limit the sample to isolated pairs based on their larger scale environment.

\subsubsection{Selection based on larger scale environment} 

\label{subsubsec:environment}
The properties of galaxy pairs depend on their environments. The SFR enhancement of interacting galaxies is influenced by the presence of another galaxy nearby and the local galaxy density \citep{Stierwalt2015,Bustamante2020}. When galaxies fall into massive halos, their cold gas content can be significantly affected by tidal interactions, ram pressure stripping, and other physical processes \citep{Wang2021, Cortese2021, Lin2023}, which ultimately change the SFRs. 

To minimize the environmental effects, we select ``isolated'' galaxy pairs of which neither has another companion within \dproj\ $=250$ kpc and \dv\ $=500~\rm km~s^{-1}$. Besides, most of the sky fields observed in the WALLABY pilot survey contain galaxy groups or clusters. To avoid spurious pairs selected due to the similarity of radial velocities of member galaxies within the groups, we further exclude galaxies within the virial radii\footnote{Galaxies in the periphery of massive groups may still exhibit different properties from galaxies in the field (e.g., through back-splashing or pre-processing). We have examined that increasing the radius of exclusion three times does not affect our primary results.} 
of massive galaxy groups (with halo mass $\log M_{200}/M_\odot \geq 13.0$, as listed in Table \ref{tab:group}) and having similar radial velocities with these groups (\dv\ $<5\sigma_{v,\rm grp}$, also see Table \ref{tab:group}) at the same time. There are three other massive groups near the WALLABY fields used in this work \citep[Virgo cluster, NGC 4261 group, NGC 5084 group;][]{Kourkchi2017}, but none of our sample galaxies is excluded due to the presence of these groups.

\begin{deluxetable*}{cCCCCCCCcc}
\tablecaption{Massive groups (clusters) targeted by WALLABY pilot survey.}
\tablehead{
\colhead{Name} & \twocolhead{Center~~~~~~} & \colhead{Distance} & \colhead{$V_{\rm hel}$} & \colhead{$M_{200}$} & \colhead{$R_{200}$} & \colhead{$\sigma_{v,\rm grp}$} & \colhead{$z_{\rm max}$} & \colhead{Area}\\
\colhead{} & \colhead{R.A.} & \colhead{Decl.} & \colhead{(Mpc)} & \colhead{($\rm km~s^{-1}$)} & \colhead{($10^{13}M_\odot$)} & \colhead{(Mpc)} & \colhead{($\rm km~s^{-1}$)} & \colhead{} & \colhead{$\rm deg^2$}
}
\label{tab:group}
\decimalcolnumbers
\startdata
Hydra I cluster & 159.174 & -27.524 & 47.5 \pm 3 & 3686 & 30.3 & 1.39 & 643 & 0.080 & 60 \\
NGC 4636 group  & 190.706 &  +2.686 & 16.2 \pm 7 & 919  &  2.5 & 0.61 & 278 & 0.080 & 60 \\
NGC 5044 group  & 198.850 & -16.390 & 27.9 \pm 5 & 2488 &  4.6 & 0.74 & 341 & 0.089 & 120
\enddata
\tablecomments{Columns: (1) names of the groups; (2)-(3) R.A. and Decl. of the center positions in J2000 \citep{Piffaretti2011}; (4) distances to the groups from \citet{Kourkchi2017}; (5) heliocentric radial velocities of the group (cluster) from \citet{Kourkchi2017}; (6)-(7) halo mass and physical radius of the groups from \cite{Reiprich2002}; (8) velocity dispersions of the groups estimated from their halo masses \citep{Evrard2008}; (9) maximum redshift probed within the WALLABY field that contains the group; (10) approximate sky area covered by the field. The NGC 4808 field has $z_{\rm max}=0.089$ and $\rm Area=30~deg^2$.}
\end{deluxetable*}

To summarize, we select galaxies with a single companion within \dproj\ $=250$ kpc and \dv\ $=500~\rm km~s^{-1}$, which should also lie beyond the virial radii of any massive groups. The final sample contains 149 \hi-detected, ``isolated'' galaxy pairs, corresponding to 298 galaxies, to which we will refer as the parent sample hereinafter. As we will describe later in Section \ref{subsec:control}, we limit our analysis to the 278 galaxies with $\log M_*/M_\odot>7.6$, which span a redshift range of $0.002<z<0.075$.

\subsection{Obtaining galaxy properties} \label{subsec:phot}

\subsubsection{Distance}
We use similar strategies as used for the ALFALFA survey \citep{Haynes2018} to derive the galaxy distances. Different methods to provide the distances are arranged in \textit{ascending} order of priority as follows:

\begin{enumerate}
    \item For galaxies with a barycentric radial velocity $v_{\rm rad}>5000~\rm km~s^{-1}$, the distance is simply estimated from the Hubble flow, with $H_0=70~\rm{km~s^{-1}~Mpc^{-1}}$. 
    \item For galaxies with a radial velocity $v_{\rm rad}\leq5000~\rm km~s^{-1}$, we use the Cosmicflows-3 Distance-Velocity Calculator\footnote{\url{http://edd.ifa.hawaii.edu/CF3calculator/}} \citep{Kourkchi2020}, which computes expectation distances based on the smoothed velocity field of the nearby Universe \citep{Graziani2019}.
    \item The redshift-independent distances for individual galaxies from the Cosmicflows-4 catalog \citep{Tully2009,Tully2023} are applied if available.
    \item If a galaxy is categorized as a member of any massive galaxy groups (clusters) with $\log M_{200}/M_\odot \geq 13.0$ in the group catalogs \citep{Tully2015, Kourkchi2017}, we assign the distance to the corresponding Brightest Group (Cluster) Galaxy.
\end{enumerate} 

In the parent sample, the number of galaxies with their distances determined by the four methods is 208, 33, 4, and 33, respectively.

\subsubsection{Masses and sizes} \label{subsubsec:mass}

We use the $g$, $r$, and $i$ band images from the DESI Legacy Imaging Surveys \citep[DR10;][]{Dey2019} to measure the magnitudes and sizes of the paired galaxies in optical bands. The official \texttt{Tractor} catalogs of DESI \citep{Lang2016} are not adopted, since their pipeline is prone to shredding extended sources at low redshifts, which are typical for our sample. We largely follow the procedures described in \citet{Wang2017} and \citet{Lin2023} for the optical photometry and measure the surface brightness profiles for each galaxy. The magnitudes are converted to the SDSS filter systems based on the calibrations of \citet{Finkbeiner2016}, \citet{Dey2019}, and formulas listed in the Legacy Surveys Data Release Description.\footnote{\url{https://www.legacysurvey.org/dr10/description/}} The $k$-corrections are applied following \citet{Chilingarian2010}.

We derive the stellar masses ($M_\ast$) of the parent sample based on the $i$ band Petrosian magnitudes\footnote{\url{http://cas.sdss.org/dr4/en/help/docs/algorithm.asp?key=mag_petro}}, with the stellar mass-to-light ratio obtained from the $g-i$ color according to \cite{Zibetti2009}.
The stellar masses of these galaxies are then converted to the halo masses $M_{\rm h}$, using the stellar mass--halo mass relation at redshift $z=0$ provided by \textsc{UniverseMachine} \citep{Behroozi2019}. 

The \hi\ masses ($M_{\rm HI}$) are derived directly from the \hi\ fluxes and distances, assuming no \hi\ self-absorption. We follow \citet{Westmeier2022} and Murugeshan et al. (in prep) to correct for the \hi\ flux discrepancies between WALLABY and ALFALFA, the latter of which is used for the control sample (Section \ref{subsec:control}). The deblending algorithm (Section \ref{sec:deblend}) is applied to obtain \hi\ fluxes when necessary. Baryonic mass and total mass are calculated through $M_{\rm b}=M_*+1.36M_{\rm HI}$ and $M_{\rm tot}=M_{\rm h}+M_{\rm b}$, respectively, where the factor 1.36 accounts for the contribution from helium and heavier elements \citep[e.g.,][]{Bigiel2010}. The radius of the \hi\ disks defined at surface density $1M_\odot\rm pc^{-2}$ ($R_{\rm HI}$) are estimated using the \hi\ size-mass relation from \citet{Wang2016}.

\subsubsection{Star formation rate} \label{subsubsec:sfr}

The SFRs of our parent sample are derived by combining the luminosity in mid-infrared (IR) and ultraviolet (UV) bands, which represent the dust attenuated and unattenuated radiation from young stars, respectively.

The UV luminosity is obtained from the Galaxy Evolution Explorer \citep[GALEX;][]{Martin2005}. We extract data from the \textsc{GUVcat} catalog \citep{Bianchi2017}, which is based on the GALEX General Release 6 and 7 (GR6+7) and includes tiles covered by the shallow All-Sky Imaging Survey (AIS) and the deeper Medium Imaging Survey (MIS). Utilizing the Kron elliptical aperture flux measured by \texttt{SExtractor} \citep{Bertin1996}, we calculate the luminosity in the FUV band. In cases where FUV data are unavailable, we resort to the NUV band photometry, encompassing 41 galaxies in the parent sample. For all the relevant UV photometry, we apply both the ``edge-of-detector'' and blending correction based on \citet{Salim2016}.

The Wide-field Infrared Survey Explorer \citep[WISE;][]{Wright2010} provides all-sky images in four mid-IR bands, W1 to W4. To measure the mid-IR luminosity, we perform aperture photometry using \texttt{Photutils} \citep{Bradley2022} with the unWISE images \citep{Lang2014} in the W3 12 $\textmu \rm m$ and W4 22 $\textmu \rm m$ bands\footnote{The effective wavelength of WISE W4 is re-calibrated by \citet{Brown2014} to be 22.8 $\textmu \rm m$, which bring the WISE W4 and IRAS 25 $\textmu \rm m$ fluxes into closer agreement.}. If the Petrosian radius in $g$ band ($R_{{\rm petro},g}$) is larger than the FWHM of the unWISE point-spread function (PSF; $6.''4$ and $12.''0$ for W3 and W4 bands, respectively), we adopt an aperture size of $3R_{{\rm petro},g}$, which we have checked to produce a consistent photometry with the Kron apertures as used in the UV bands. Otherwise, we adopt the PSF-fitting photometry from the AllWISE catalog \citep{Wright2010}, which perform better than the aperture photometry for these sources since the W4 images are relatively shallow \citep[e.g.,][]{Li2023b}. Corrections are applied to the aperture photometry measurements based on instructions developed by \citet{Jarrett2013} and described in the Explanatory Supplement to the WISE Preliminary Data Release Products\footnote{\url{https://wise2.ipac.caltech.edu/docs/release/prelim/expsup/wise_prelrel_toc.html}} \citep[see also][]{Wright2010, Jarrett2011}.  With the FUV (NUV) + W4 band luminosity, we calculate the SFRs using the relations tabulated in \citet{Kennicutt2012}, which enable us to probe the SFRs on timescales short enough to capture the evolution of interacting stages \citep{Davies2015}: 
\begin{eqnarray}
\begin{array}{ll}
    {\log\rm SFR} &= \log (L_{\rm FUV} + 3.89L_{25\textmu \rm m}) - 43.35\\
    {\log\rm SFR} &= \log (L_{\rm NUV} + 2.26L_{25\textmu \rm m}) - 43.17
\end{array}\label{eq:sfr}
\end{eqnarray}
where the units of SFR and luminosity $L$ are $M_\odot~\rm yr^{-1}$ and $\rm erg~s^{-1}$, respectively. The luminosity in the WISE W4 band is used as a proxy of $L_{25\textmu\rm m}$, since the flux ratios between these two bands are expected to vary only in a narrow range, from 0.9 for late-type galaxies to 1.1 for early-type galaxies \citep{Jarrett2013}. 

In Appendix \ref{app:control}, we validate our SFR measuring procedure by comparing its outcomes with the SFRs obtained from spectral energy distribution (SED) fitting. The latter has been used to derive the SFRs for our control sample (defined in Section \ref{subsec:control}).

\subsection{The control sample} \label{subsec:control}

\begin{figure*}
    \centering
    \includegraphics[width=0.9\textwidth]{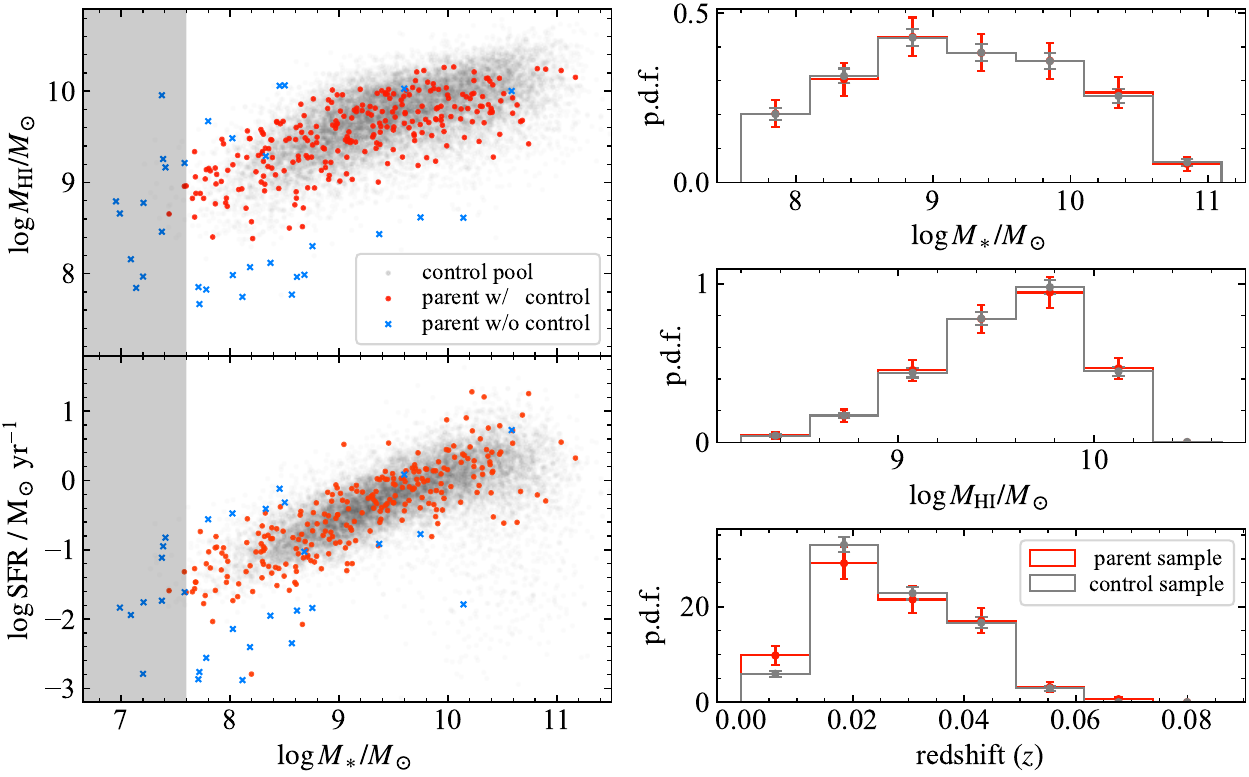}
    \caption{An overview of the parent sample and the control sample. \textbf{Left}: The distribution of galaxies in the \hi-stellar mass plane (upper) and the SFR-stellar mass plane (lower). Gray, red, and blue symbols represent galaxies in the control pool, paired galaxies with at least 5 control galaxies, and paired galaxies omitted due to the lacking of control galaxies, respectively. The shaded regions correspond to $\log M_\ast/M_\odot<7.6$, where all the galaxies are excluded from analysis. \textbf{Right}: The stellar mass ($M_\ast$), \hi\ mass ($M_{\rm HI}$), and redshift ($z$) distributions of the parent sample with $\log M_\ast/M_\odot \geq 7.6$ (red histograms) and the control sample (gray histograms).}
    \label{fig:sample}
\end{figure*}

We compare the galaxies in our parent sample with a control sample,  aiming to explore the effects of galaxy interactions on their SFRs at fixed stellar mass and atomic gas content.

The control pool is constructed through cross-matching\footnote{The search radius is $3''$, with a tolerance of 250 $\rm km~s^{-1}$ in the radial velocities. Coordinates of the optical counterparts in ALFALFA are used during the matching.} the GALEX-SDSS-WISE Legacy Catalog \citep[GSWLC;][]{Salim2016,Salim2018} with the extragalactic \hi\ source catalog from ALFALFA \citep{Haynes2018}, resulting in 15094 galaxies with \hi\ mass measurements. The SFR is obtained from the SED fitting of \citet{Salim2018}, which is consistent with our measurements as shown in Appendix \ref{app:control}. Using the same strategy applied to our parent sample, we re-calculate the stellar masses for the control pool based on \citet{Zibetti2009}, to eliminate the systematics between the two samples. We employ the optical photometry data from the NASA-Sloan-Atlas \citep{Blanton2011}, specifically the elliptical Petrosian magnitudes, for the derivation of stellar masses, thus slightly reducing the number of galaxies in the control pool to 14449. No isolation criterion is applied to the control pool, as we aim to compare interacting galaxies with general galaxy populations at the same $M_\ast$ and $M_{\rm HI}$. Furthermore, the incidence of galaxies in massive groups, where the WALLABY pilot survey is intentionally targeting, is low for general galaxy samples, especially for the \hi-rich ones detected in ALFALFA (see also Appendix \ref{app:env}).

For each galaxy in the parent sample, we search for galaxies in the control pool with similar stellar and \hi\ masses, as well as similar redshifts: 
\begin{equation}
    \delta\log M_*<0.2,~\delta\log M_{\rm HI}<0.2,~\delta z<0.01. \label{eq:control}
\end{equation}
For the galaxies matched through Equation \ref{eq:control}, we then select the five ``nearest'' galaxies in the parameter space based on the parametric distance $D$ defined as follows:
\begin{equation}
    D^2=\left(\frac{\delta\log M_*}{0.1}\right)^2+\left(\frac{\delta\log M_{\rm HI}}{0.1}\right)^2+\left(\frac{\delta z}{0.01}\right)^2. \label{eq:d_param}
\end{equation}
Therefore, the control sample comprises only a portion of the galaxies in the control pool. We have examined that our results remain unaffected when altering any of the normalization factors on the denominator in Equation \ref{eq:d_param} by a factor of two.

Among the 298 galaxies in the isolated galaxy pairs, 41 cannot be matched with five control galaxies mainly due to their low masses. Thus, we further limit our sample to galaxies with $\log M_*/M_\odot > 7.6$, where the fraction of galaxies with controls exceeds 90\%, leaving 278 galaxies in the parent sample (Section \ref{subsubsec:environment}). After applying the mass cut, there remain 21 (8\%) galaxies with fewer than 5 control galaxies, and we omit them in the following analyses. Properties of the parent and control sample are summarized in Figure \ref{fig:sample}.

% ------------------------
\section{Deblending Gas-rich Galaxy Pairs in \hi\ Data Cubes} \label{sec:deblend}

As mentioned in Section \ref{subsec:wallaby}, blending of the \hi\ sources does occur in WALLABY data products and it hinders us from getting the \hi\ properties of the individual galaxies. At the redshift of our sample ($z\lesssim0.06$), there are $\sim 4\%$ of galaxies suffering from blending in the WALLABY fields used in this work, which is acceptable for general research. However, the blending problem becomes inevitable when we focus on galaxy pairs with $d_{\rm proj}<100$ kpc, where significant SFR enhancement is found in previous works \citep[e.g.,][]{Ellison2008,Patton2013,Stierwalt2015,Pan2019}. Among the 278 galaxies in our parent sample, 116 (42\%) are blended in terms of \hi\ detections, highlighting the necessity to process these data separately. 

The blending problem could be partially solved by tuning the parameters of the source detection algorithms like \texttt{SoFiA}, but this will inevitably increase the rate of spurious detection for well resolved \hi\ sources. Besides, the upcoming data released by WALLABY and future \hi\ surveys will always encounter the same problem when pushing the utilization of the data toward the limit of its spatial resolution.

Based on the considerations above, we develop an algorithm\footnote{\url{https://github.com/BetaGem/wallaby-galaxy-pair}} to separate the fluxes in blended \hi\ detections. It should be emphasized that the algorithm presented here only segments the voxels in the data cubes into separate parts. Dividing the fluxes within individual voxels (e.g., through data modeling) is beyond the scope of this work. Our primary purpose is to provide an automatic and mathematical deblending method for the data, rather than performing this task manually and subjectively. 

\begin{figure*}
    \centering
    \includegraphics[width=\textwidth]{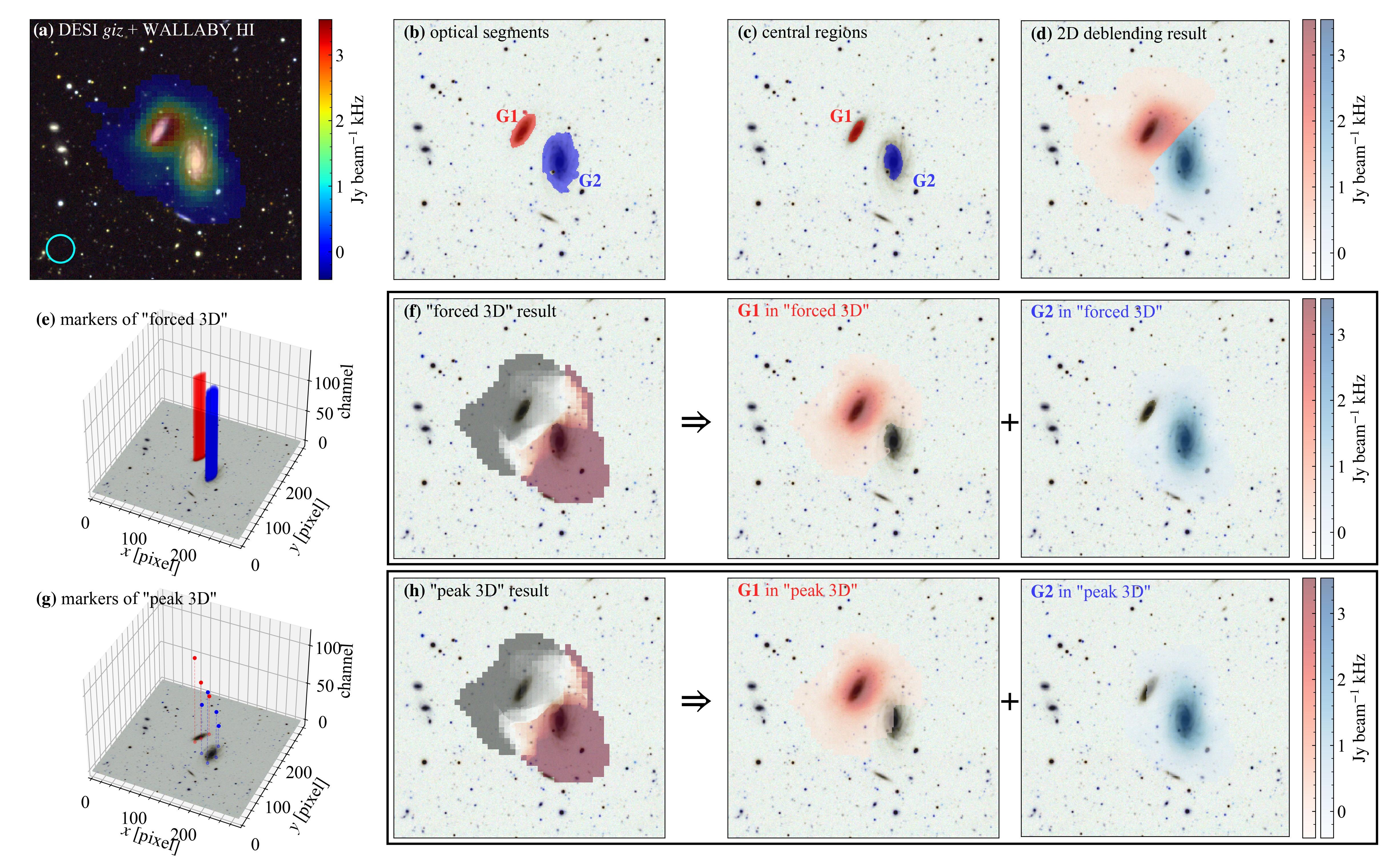}
    \caption{Illustration of the deblending algorithm for a pair of disk galaxies. \textbf{(a)} Optical image from the DESI Legacy Imaging Survey, overlaid with moment-0 map from WALLABY. The cyan circle in the bottom left represents the synthesized beam of WALLABY. \textbf{(b)} Results of source detection and deblending on the optical image, with galaxies G1 and G2 labeled in red and blue, respectively. \textbf{(c)} ``Central regions'' of the galaxies in the optical image used to establish initial markers in Steps 2 and 3. \textbf{(d)} 2D deblending result from Step 2, with \hi\ fluxes associated with galaxies G1 and G2 distinguished in color. \textbf{(e)} 3D illustration of the initial markers for the ``forced 3D'' method. The optical image is attached for reference. \textbf{(f)} Deblending result of the ``forced 3D'' method, using markers shown in panel (e). Among the three plots outlined by the black rectangle, the leftmost one is color-coded by the fraction of channels assigned to a galaxy at a given spaxel, ranging from 0\% (gray), 50\% (white) to 100\% (dark red) for galaxy G1. The two plots on the right show the deblended moment-0 maps of the two galaxies. \textbf{(g)} 3D illustration of the markers for the ``peak 3D'' method. Data points represent \hi\ peaks in the data cube, with the dashed lines indicating their 2D projections on the sky. \textbf{(h)} Similar with panel (f), but using the ``peak 3D'' method.}
    \label{fig:deblend}
\end{figure*}

\subsection{The deblending algorithm}

The basic idea of our algorithm is to perform segmentation in three-dimensional (3D) \hi\ data cubes, leveraging optical information as priors. We utilize the \texttt{watershed} algorithm implemented in \texttt{skimage} \citep{vanderWalt2014} to deblend the \hi\ data. \texttt{watershed} is a traditional algorithm for image segmentation and has been widely used on 2D images. It begins by taking user-defined \textit{markers} as starting points and viewing pixel values as a local topography. The basins are then flooded from the markers until the basins associated with different markers meet on watershed lines. For our purposes, we apply the algorithm to 3D \hi\ data cubes, with the expectation that the inclusion of a dimension of frequency will improve the deblending results. The procedures are described step-by-step as follows, and an example is shown in Figure \ref{fig:deblend} to illustrate the deblending algorithm. A 3D visualization of the deblending results using \texttt{SlicerAstro} \citep{Punzo2017} is provided in Appendix \ref{app:3D}. A recent application of the algorithm can be found in \citet{Wang2023a}.

\vspace{10pt}

\noindent\textbf{Step 1. Markers based on optical images.}
As the preparation step, we perform source detection and segmentation on optical images using \texttt{Photutils} \citep[the \texttt{Astropy} package for photometry;][]{Bradley2022}, with star masks generated according to the Gaia DR3 catalog \citep{GaiaCollaboration2023}. This provides us with the positions and spatial extent of galaxy pairs in the optical images (Figure \ref{fig:deblend}b). After the segmentation, the optical images are smoothed by a Gaussian kernel with an FWHM of $0.''8$ (3 pixels). Then within each optical segment, we select pixels with flux higher than the segment average to define the ``central regions'' of galaxies as shown in Figure \ref{fig:deblend}c, which will be used as initial markers for the \texttt{watershed} algorithm.

\noindent\textbf{Step 2. Two-dimensional segmentation.} Before applying our algorithm to 3D data cubes, we run the 2D deblending algorithm first for illustration and comparison. In this step, \texttt{watershed} is applied on moment-0 maps of the \hi\ data, with the optical segmentation maps used as initial markers. To avoid producing spurious \hi\ segments, we require a minimum growth in the basin area of 5 pixels, otherwise the marker corresponding to that basin will be removed and the \texttt{watershed} will be run again. For galaxy pairs with large separations, the deblending results obtained here are almost identical to those of the 3D methods described below.

\noindent\textbf{Step 3. The ``forced 3D'' method.} Under the assumption that the spatial distribution of \hi\ is roughly associated with the optical counterparts, we set the initial markers as the spaxels covered by those optical central regions for every channel of the data cubes (i.e., all the channels have the same markers). Then the data cube is smoothed along the frequency axis with a Gaussian kernel ($\sigma=4$ channels) to increase the signal-to-noise ratio and to avoid unphysical results. Finally, \texttt{watershed} is performed on the entire data cube (not channel-wise) with these markers. A threshold of minimum growth in the basin area is also set as in Step 2. Since the markers appear to anchor the data cubes in this step, we term the method ``forced 3D.''

\noindent\textbf{Step 4. The ``peak 3D'' method.} In Step 3, the 3D information is not fully utilized since the markers are essentially 2D. So we try to make further improvement by setting local maximum (\hi\ peaks) in the data cube as the initial markers for \texttt{watershed}. The peaks are detected using our modified version of \texttt{find\_peaks},\footnote{\url{https://photutils.readthedocs.io/en/stable/api/photutils.detection.find_peaks.html}} originally programmed for 2D images. Labels of these \hi\ peaks are assigned consistently with the deblending result in Step 3. \texttt{watershed} is then run on the data cube with these \hi\ peaks as the markers. 

From the description above, we see that the algorithm is based on two assumptions. First, the optical counterparts of the galaxy pairs should be separable in optical images. Second, the \hi\ contents are roughly associated with the corresponding optical counterparts spatially. This means the algorithm is not applicable to galaxies with heavily distorted or highly irregular \hi\ morphology, such as some of the early-type galaxies \citep{Serra2012} and mergers in the final stage. 

We perform mock tests in Appendix \ref{app:deblend} to quantify the uncertainties associated with our deblending techniques, utilizing non-interacting galaxies in WALLABY that do not satisfy the criteria on projected distances and radial velocity offsets (Section \ref{subsubsec:d&v}). To summarize, the tests indicate that the uncertainties in the deblended \hi\ fluxes barely exceed the errors attributed to the observational noise. The ``peak 3D'' method performs the best among all these methods in most cases, and the 2D method generally performs worse than its 3D counterparts. Interestingly, the ``forced 3D'' method shows the lowest uncertainties in scenarios where none of the methods can deblend the data cubes accurately, as shown in Figure \ref{fig:deblend_error}. For blended pairs in our parent sample, we derive the \hi\ masses from the outputs of Step 4. In three cases, the ``peak 3D'' method fails due to an advanced merging stage, and we resort to the output of the ``forced 3D'' method from Step 3. We have checked that excluding these galaxies does not affect our results.

% ------------------------
\section{Star formation enhancement in galaxy pairs} \label{sec:results}
We can now investigate how the interactions affect the SFR of \hi-rich galaxy pairs. For each galaxy in our parent sample, its SFR enhancement (\dsfr; SFR suppression for negative values) is defined as the SFR offset between itself and the median ($\textmu_{1/2}$) of the control galaxies:
\begin{equation}
    \rm \Delta\log SFR=\log SFR_{pair} - \textmu_{1/2}\big(\log SFR_{control}\big). \label{eq:dsfr}
\end{equation}
Since we have controlled the \hi\ mass, \dsfr\ will vary in any scenarios where SFR does not change proportionately with \hi\ mass (i.e., both altering the SFR at fixed \hi\ mass and changing the \hi\ mass at fixed SFR will cause the variations in \dsfr).

In the following of this section, we will summarize the dependence of \dsfr\ on several properties of galaxy pairs (Section \ref{subsec:sfr_depend}), and reveal the difference between galaxies with different stellar masses (Section \ref{subsec:2mass}). Implications of these results and physical processes involved in the interactions are discussed in Section \ref{sec:discuss}.

\subsection{Dependence on basic properties of galaxy pairs} \label{subsec:sfr_depend}

\begin{figure*}
    \centering
    \includegraphics[width=\textwidth]{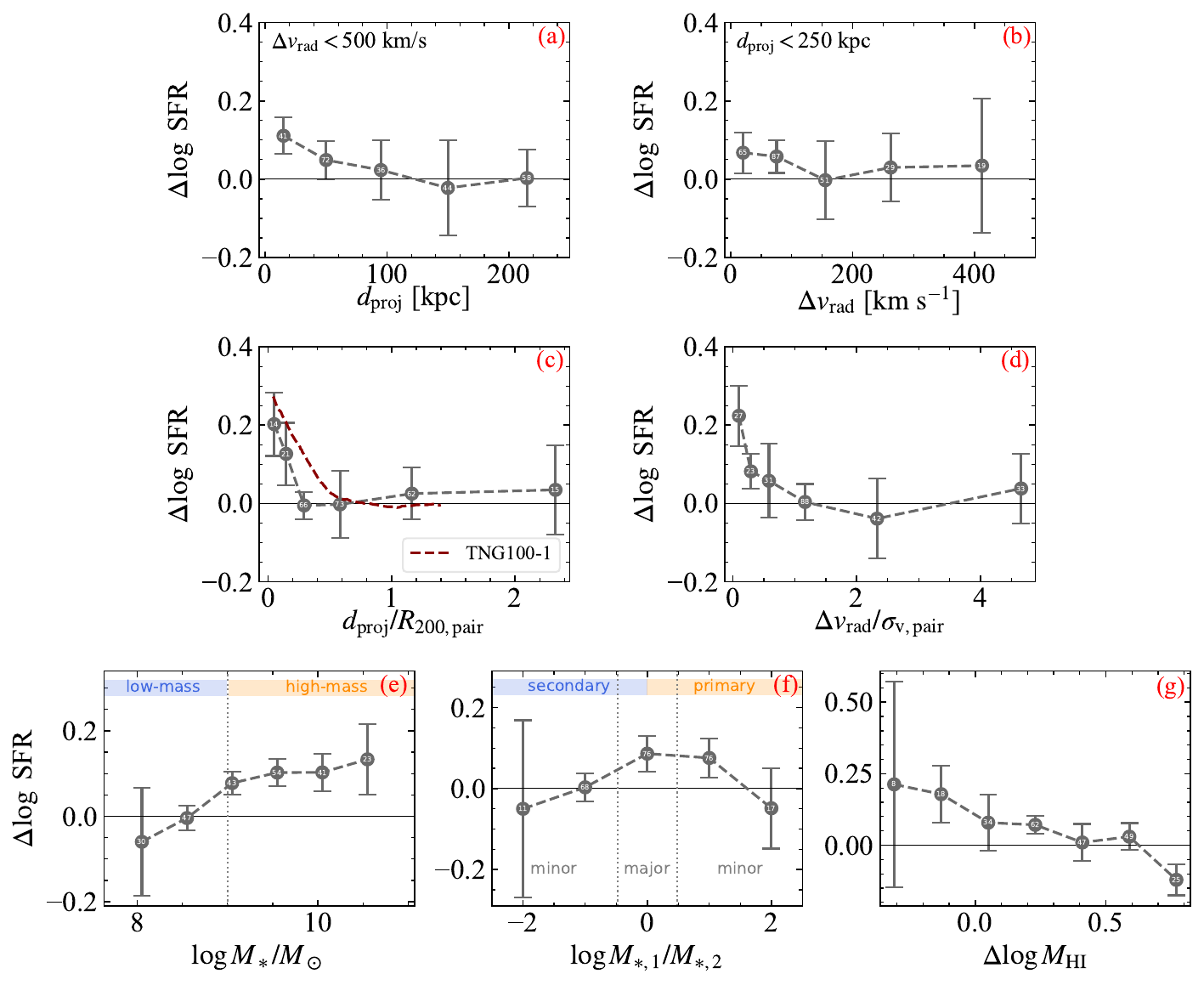}
    \caption{Dependence of the median SFR enhancement on the basic properties of galaxy pairs in the parent sample: \textbf{(a)} projected distance; \textbf{(b)} radial velocity offset; \textbf{(c)} normalized projected distance; \textbf{(d)} normalized radial velocity offset; \textbf{(e)} stellar mass; \textbf{(f)} stellar mass ratio; \textbf{(g)} \hi-richness. Data points are placed at the center of each bin, overlaid with the number of galaxies in the bins. Error bars represent the $1\sigma$ uncertainty of the medians, estimated from $10^4$ iterations of bootstrapping. The black horizontal lines correspond to no SFR enhancement (\dsfr\ $=0$). The brown dashed line in panel (b) represents the result from the TNG100-1 run of IllustrisTNG \citep{Patton2020}. In panels (e) and (f), the blue and orange labels illustrate the definition of subsamples. The vertical dotted lines mark the division between low- and high-mass galaxies in panel (e), or stellar mass ratios of 1:3 and 3:1 in panel (f). Bins with fewer than five galaxies are not shown in the figure.}
    \label{fig:sfr_depend}
\end{figure*}

\textbf{Pair separations.}
In previous studies, the projected distance \dproj\ and radial velocity offsets \dv\ have been common metrics to demonstrate the elevated SFRs during galaxy interactions \citep{Ellison2008, Patton2013, Patton2020, Li2023a}. We begin with this straightforward exercise with our sample, shown in Figure \ref{fig:sfr_depend}a.\footnote{From Figure \ref{fig:sfr_depend} to \ref{fig:high-mass}, both galaxies of each pair contribute separately to the plots.}
As expected, there is a discernible rise in \dsfr\ as \dproj\ decreases, with significant positive values occurring at \dproj\ $\lesssim$ 100 kpc, consistent with previous observations \citep[e.g.,][]{Patton2013, Bustamante2020}. Along the line of sight, the separations of galaxy pairs can be quantified by the radial velocity offset \dv. On average, a smaller value of \dv\ signifies a later stage of galaxy mergers or slower encounters, so an anti-correlation between \dsfr\ and \dv\ is expected. We plot the median SFR enhancement against \dv\ in Figure \ref{fig:sfr_depend}b. Indeed, the SFR enhancement is observed only at \dv\ $\lesssim 100~\rm km~s^{-1}$, but the anti-correlation is not significant.

From first principles, however, it is not even-handed to compare \dproj\ and \dv\ for galaxies with distinct stellar mass \citep[e.g.,][]{Park2009}, especially when it covers over three orders of magnitude for our sample. To normalize the pair separations with respect to the masses, we approach the dynamics of the pairs through the classical two-body problem, which regards the system as a satellite orbiting an immobile central ``reduced galaxy.'' Its halo mass, $M_{\rm 200,pair}$, is given by the sum of the halo masses of galaxy pairs (Section \ref{subsubsec:mass}). Then, its virial radius ($R_{\rm 200,pair}$) and velocity dispersion ($\sigma_{v,\rm pair}$) are calculated based on the halo mass, following equations in \citet{Cimatti2020} and \citet{Evrard2008}, respectively. Finally, the normalized separations of galaxy pairs are given by \dprojn\ and \dvn.

In panels (c) and (d) of Figure \ref{fig:sfr_depend}, we show similar correlations as in panels (a) and (b), except that the horizontal axes are normalized by the corresponding ``reduced'' halo properties. Instead of maintaining a low level of $\sim$ 0.1 dex, \dsfr\ increases rapidly as the normalized versions of \dproj\ and \dv\ fall below unity and approach zero, reaching a maximum of $\gtrsim 0.2$ dex, along with higher statistical significance. The presence of a critical value of order unity in both panels suggests that some physical processes begin to take effect at these scales. The trend against \dprojn\ broadly agrees with previous observations \citep{Park2009} and the IllustrisTNG simulation \citep{Patton2020}, although different sample selection criteria could weaken the significance of the comparisons.\footnote{Although \citet{Patton2020} normalized the 3D distance of galaxy pairs with the sum of their halo radius, both the denominator and numerator in their Equation 4 are $\sim$ 1.5 times larger than in our definition, making the ratios comparable in a statistical sense.}

\textbf{Stellar mass.}
In Figure \ref{fig:sfr_depend}e, we see that the median \dsfr\ changes little around 0.1 dex for galaxies with $\log M_\ast/M_\odot > 9$ (``high-mass galaxies'' hereinafter). Low-mass galaxies with $\log M_\ast/M_\odot < 9$, however, display no enhancement in their SFRs, seemingly in contrast to general expectations \citep[e.g.,][]{Bekki2008, Kado-Fong2020, Subramanian2024}.
Since galaxies in massive galaxy groups have been omitted from our sample, the environmental effects of massive groups do not play a role. Our selection of isolated galaxy pairs further weakens the impact of nearby massive halos \citep{Stierwalt2015}. These analyses indicate that interactions within the galaxy pairs are the likely origin of this phenomenon, as we will further discuss in Section \ref{subsec:low-mass}.

\textbf{Stellar mass ratio.} Figure \ref{fig:sfr_depend}f illustrates how \dsfr\ varies with the stellar mass ratio $\log M_{\ast,1}/M_{\ast,2}$. Positive values of $\log M_{\ast,1}/M_{\ast,2}$ suggest that the target galaxy is more massive than its companion, and vice versa. Tentatively, galaxies with major companions ($1/3<\log M_{\ast,1}/M_{\ast,2}<3$) display the highest \dsfr\ among all the mass ratios, in agreement with previous studies \citep[e.g.,][]{Ellison2008,Hani2020,Bottrell2023}. But the low level of \dsfr\ for the major mergers ($0.09\pm0.04$ dex) suggests that stellar mass ratio may not be the primary driver of a substantial SFR enhancement for our sample. From now on, we will refer to the galaxies with $\log M_{\ast,1}/M_{\ast,2}>0$ ($<0$) as primary (secondary) galaxies.

\textbf{HI-richness.} 
As the cold gas reservoir of star formation, \hi\ dominates the mass budget of cold interstellar medium (ISM) and is strongly correlated with the specific SFR of galaxies \citep{Saintonge2017, Saintonge2022}. Figure \ref{fig:sfr_depend}g displays the correlation between SFR enhancement and \hi-richness ($\Delta \log M_{\rm HI}$) of interacting galaxies for our parent sample. The \hi-richness is defined as the difference between actual and expected \hi\ mass based on the \hi-stellar mass relation for star-forming galaxies at $z\approx0$ \citep{Guo2021}:\footnote{Employing the equation 4 of \citet{Janowiecki2020} based on the xGASS sample gives qualitatively consistent results. Notice that the lower-case $\delta \log M_{\rm HI}$ has a different definition in Equation \ref{eq:control} and \ref{eq:d_param}, where it represents the \hi\ mass difference between two galaxies.}
\begin{equation}
    \Delta \log M_{\rm HI}=\log M_{\rm HI} - 0.42\log M_\ast - 5.35. \label{eq:dMHI}
\end{equation}
We remind the reader that the \hi\ mass is controlled when evaluating the SFR enhancement.

As indicated by the gray line, the most \hi-poor galaxies in our sample manifest the highest level of SFR enhancement, reaching $\gtrsim 0.2$ dex at $\Delta \log M_{\rm HI}= -0.3$, in good agreement with the binary merger simulations carried out by \citet{Scudder2015}. The negative correlation between \dsfr\ and gas-richness is also consistent with the literature \citep{Scudder2015, Cao2016, Garay-Solis2023}. For galaxies with elevated \hi-richness, the median \dsfr\ fluctuates around zero, showing less significant SFR enhancement as the galaxy pairs at higher redshift \citep{Patton2020, Shah2022}. Simulations expect very gas-rich mergers to reach \dsfr\ $\gtrsim 1$ dex during their final coalescence \citep{Scudder2015, Moreno2019}. These interacting galaxies are often treated as a whole system in observations \citep{Ellison2018, Pan2019}, and are beyond the scope of this paper due to their inseparable \hi.

\subsection{Differences between low- and high-mass paired galaxies} \label{subsec:2mass}

As hinted in the last section, \dsfr\ of interacting galaxies seems to behave differently in the low-mass regime (Figure \ref{fig:sfr_depend}e). We split our sample according to their stellar mass at $\log M_\ast/M_\odot=9$ to further investigate the SFR enhancement characteristics.

\begin{figure}
    \centering
    \includegraphics[width=0.46\textwidth]{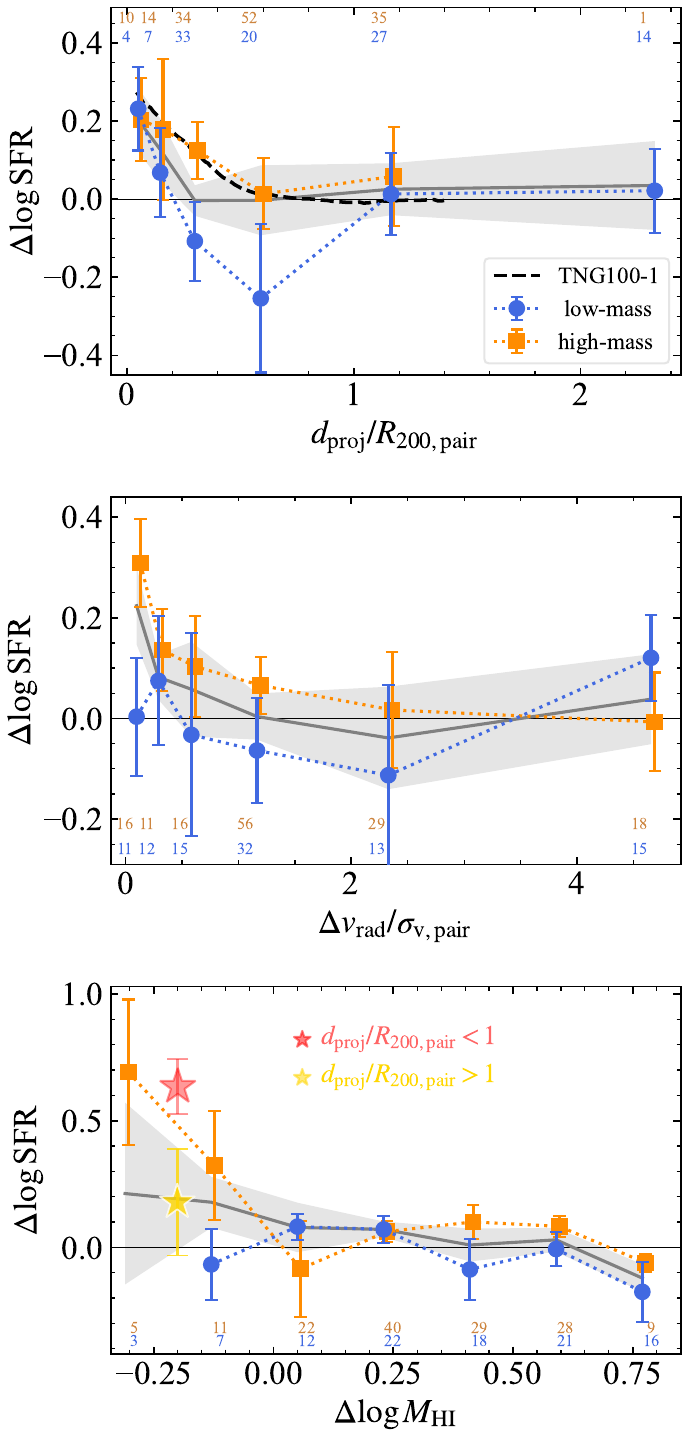}
    \caption{SFR enhancement as a function of \dprojn\ (top), \dvn\ (middle), and \hi-richness (bottom). Blue and orange lines stand for the low- and high-mass galaxies, respectively, with a small horizontal shift applied for clarity. The solid gray lines represent trends for the parent sample as shown in Figure \ref{fig:sfr_depend}, with the shaded region indicating the uncertainties. The red (yellow) star in the bottom panel stands for high-mass galaxies with $\Delta \log M_{\rm HI}<0$ and \dprojn\ $<1$ ($>1$). The numbers at the top or bottom represent the number of galaxies in each bin.}
    \label{fig:high_low_mass}
\end{figure}

The top panel of Figure \ref{fig:high_low_mass} reveals dissimilar behavior between low- and high-mass galaxies. Despite some fluctuation within the errors, the SFR enhancement of high-mass galaxies is initiated as the halos of galaxy pairs come into contact at \dprojn\ $\lesssim 1$. Meanwhile, the SFR of low-mass galaxies is marginally suppressed at this stage, with minimal \dsfr\ $=-0.25 \pm 0.18$ dex ($1.4\sigma$). It only rises again at \dprojn\ $\lesssim 0.4$ and finally catches up with the high-mass galaxies at smaller distances of \dprojn\ $\lesssim 0.2$. As for the radial velocity offset \dv, the middle panel of Figure \ref{fig:high_low_mass} illustrates that whereas the SFR enhancement for high-mass galaxies rises with decreasing \dv$/\sigma_{\rm v,pair}$, low-mass galaxies show no significant variation. Segregating the parent sample into primary and secondary galaxies shows that both follow the trends of the parent sample within the uncertainties (not shown in the figures). This suggests that the distinct behavior of low-mass galaxies does not originate from their identity as secondary galaxies in the pairs, and that the absolute values of stellar mass do play a role. 

From the bottom panel of Figure \ref{fig:high_low_mass}, it becomes clear that the \hi-poor galaxies with elevated SFR enhancement lie in the high-mass regime, where the simulated galaxies of \citet{Scudder2015} are also located. Specifically, when the \hi-richness falls below $\Delta \log M_{\rm HI}\approx 0$, the \dsfr\ of high-mass galaxies increases rapidly. On the other hand, low-mass galaxies do not exhibit the same pattern. Separating the high-mass galaxies with different projected distances, we can see that the SFR enhancement in relatively \hi-poor high-mass galaxies mainly happens at \dprojn\ $<1$, signifying an effective star formation triggering in these galaxies through external disturbances.\footnote{While the most vigorous starbursts often show signatures of \hi\ absorption against the radio continuum, the 1.4 GHz continuum brightness of these \hi-poor high-mass galaxies, as obtained from the NRAO VLA Sky Survey \citep{Condon1998} and averaged over the WALLABY beam, is mostly below 10 K. This is much lower than the average spin temperature of \hi, suggesting that the \hi\ mass measurement is not significantly affected by unresolved \hi\ absorption lines \citep{Dickey1990}.}
Visual inspection of these galaxies reveals compact morphology in unWISE W4 band images, implying efficient gas inflows toward galactic centers, as has been discussed in previous works \citep{Keel1985, Kewley2006, Pan2019, Thorp2019, He2024}. A detailed study on the spatially resolved SFR enhancement for our sample, though, is beyond the scope of this paper.

% ---------------------------
\section{Discussion} \label{sec:discuss}

\subsection{Dependence on the merging stage} \label{subsec:psd}

\begin{figure}
    \centering
    \includegraphics[width=0.46\textwidth]{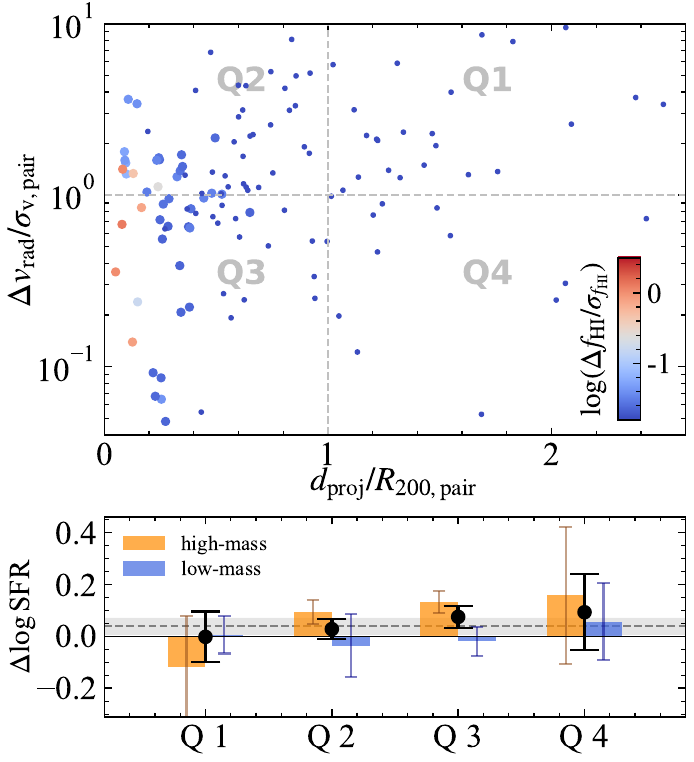}
    \caption{\textbf{Top}: distribution of galaxy pairs in the PSD. Color-coding represents the mean deblending error of \hi\ mass (see Appendix \ref{app:deblend} for details), with blended pairs highlighted by large data points. The dashed lines divide the plane into quadrants Q1 to Q4. \textbf{Bottom}: \dsfr\ of galaxy pairs in different quadrants (black points). The high- and low-mass galaxies are shown in orange and blue, respectively. The median \dsfr\ of the parent sample and its uncertainty are depicted with the gray dashed line.}
    \label{fig:psd}
\end{figure}

Before diving into the detailed physical mechanisms, we first characterize the evolutionary stages of galaxy pairs with the projected phase-space diagram (PSD; Figure \ref{fig:psd}), where normalized radial velocity offsets are plotted against projected distances. The PSD is divided into four quadrants (Q1 to Q4) with \dprojn\ $=1$ and \dvn\ $=1$ to separate galaxy pairs into different interaction stages. Q1 mainly includes galaxy flybys or early-stage mergers, where the interaction remains the weakest. Q3 includes more late-stage mergers, while Q2 and Q4 prefer galaxy pairs near the pericenters and apocenters of their orbits, respectively.

The black points in the bottom panel show how \dsfr\ changes with quadrants. The SFR enhancement is consistent with zero in Q1, while all the remaining quadrants have enhanced SFRs, with Q3 being the most significant ($0.08\pm0.04$ dex). The trend is consistent with the results of Figure \ref{fig:sfr_depend}c and \ref{fig:sfr_depend}d, but more clearly shows the independent effects of spatial and velocity offsets, and physically combines them into a temporal sequence. Within this sequence, the \hi-selected galaxy pairs in our sample will be biased toward early-stage interactions, as galaxy pairs experiencing multiple pericenter passages tend to show pronounced \hi\ disturbances or coalescence morphology \citep[e.g.,][]{Feng2020, Engler2023}, thus mitigating the orbital complexity. Such a possible sequence provides us the basis for discussing the effects by quantifying the strength of physical processes that depend on positions in the PSD. 

Separating the sample into high- and low-mass galaxies again reveals differences between the two groups. Particularly in Q3, where the interaction is the strongest, high-mass galaxies exhibit an enhanced SFR, while low-mass galaxies do not. This is qualitatively consistent with the results in Figure \ref{fig:sfr_depend}e. However, the large error bars hinder us from further splitting our sample in the PSD.

\subsection{Quantifying hydrodynamic and gravitational effects} \label{subsec:hydro_tidal}

The interactions between galaxies can be broadly categorized into two classes from first principles: hydrodynamic effects and tidal force. Numerical simulations show that both kinds of interactions can drive gas inflows \citep{Barnes1991,Blumenthal2018}, as well as lead to gas compression, producing dense gas that fuels star formation \citep{Moreno2019, Sparre2022, Petersson2022}. 

We parameterize the strength of cold gas collision by dividing the projected distance of galaxy pairs with the sum of their \hi\ disk radii ($R_{\rm HI}$; Section \ref{subsubsec:mass}):
\begin{equation}
    r_{\rm gas}=\frac{d_{\rm proj}}{R_{\rm HI,1} + R_{\rm HI,2}}, \label{eq:R_def}
\end{equation}
where the subscripts 1 and 2 represent the two galaxies involved. In terms of definition, \rgas\ is similar to the normalized separation \dprojn\ we defined in Section \ref{subsec:sfr_depend} but is normalized with the properties of gas disks rather than the dark matter halos. From the gray line in Figure \ref{fig:low-mass}b, it is clear that \dsfr\ increases with decreasing \rgas, which is expected, since \rgas\ is strongly correlated with \dprojn, with a Pearson $r$-value of 0.9. However, a crucial difference between these two parameters is their physical implications, as we will discuss later.

As for tidal interactions, we calculate the dimensionless tidal parameter for each galaxy to quantify the instantaneous tidal perturbation exerted by a companion. Following \cite{Wang2022} and \citet{Lin2023}, we have
\begin{equation}
    S_{\rm tid}=\left(\frac{M_{\rm c}}{M_{\rm t}}\right)\left(\frac{R_{\rm25}}{d_{\rm proj}}\right)^2\left(\frac{V_{\rm circ}}{\sqrt{(\Delta v_{\rm rad})^2+V_{\rm circ}^2}}\right), \label{eq:S_def}
\end{equation}
where $M_{\rm t}$ and $M_{\rm c}$ stand for the total masses (sum of baryonic and dark matter halo masses; Section \ref{subsubsec:mass}) of the target galaxy and the companion, respectively. $R_{\rm 25}$ is the isophotal radius measured at $25~{\rm mag~arcsec^{-2}}$ in $g$ band. $V_{\rm circ}$ represents the rotational velocity of the target galaxy, estimated from the baryonic Tully-Fisher relation \citep{McGaugh2000}. For each galaxy in our sample, we only consider the tidal perturbation from its closest companion. Since the galaxies we select are isolated pairs, $S_{\rm tid}$ can well represent the total strength of external tidal perturbations.

In the upcoming sections, we will discuss the impact of various physical processes on the SFR enhancement of galaxy pairs, using the parametrization above to aid in our analyses. The disparity between low- and high-mass galaxies exhibited in Figure \ref{fig:high_low_mass} indicates a possible difference in their evolutionary paths, motivating us to treat them separately.

\subsection{Evolution of low-mass paired galaxies} \label{subsec:low-mass}

\begin{figure}
    \centering
    \includegraphics[width=0.46\textwidth]{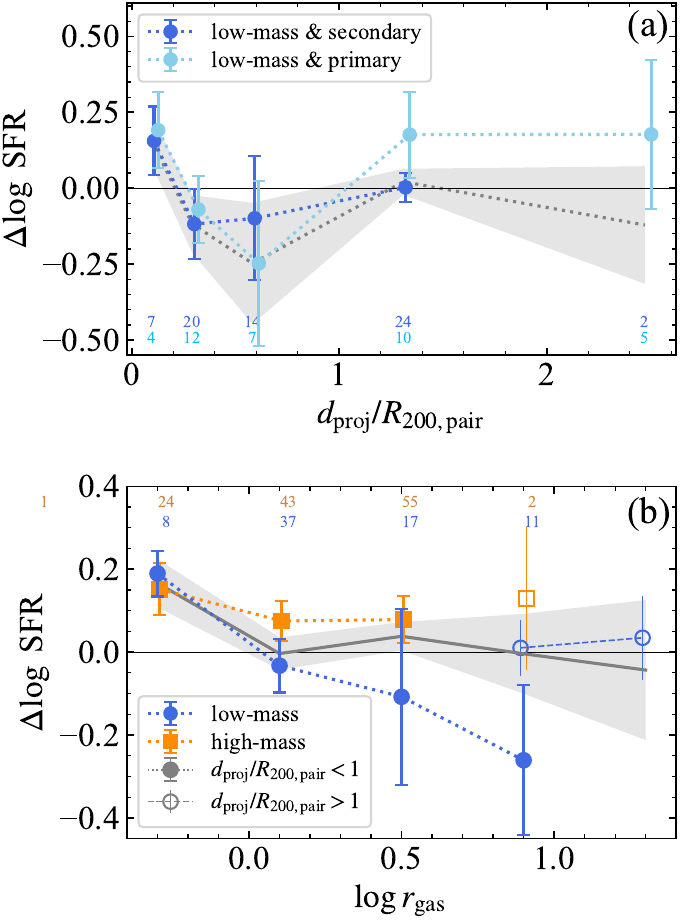}
    \caption{\textbf{Top}: SFR enhancement as a function of \dprojn\ for low-mass galaxies. The gray dotted line represents the low-mass galaxy sample, as shown in the top panel of Figure \ref{fig:high_low_mass}. The dark and light blue lines show the correlations for low-mass secondary and low-mass primary galaxies, respectively. \textbf{Bottom}: SFR enhancement as a function of \rgas. The gray line represents the entire parent sample, which is subdivided into four groups according to the stellar mass and \dprojn. The numbers at the top or bottom correspond to the solid data points.}
    \label{fig:low-mass}
\end{figure}

Low-mass galaxies have a shallow gravitational potential and a high gas fraction, and are thus more susceptible to external perturbations. Meanwhile, they are generally more metal-poor and have lower surface densities, where \hi\ has closer connections to star formation \citep{Bigiel2010, Bacchini2019, Hunter2024}. Complexities of the low-mass galaxy evolution have been glimpsed from the correlation between \dsfr\ and \dprojn\ in Figure \ref{fig:high_low_mass}. It shows that low-mass galaxies exhibit a non-monotonic trend with \dprojn, with marginal SFR suppression at $1.4\sigma$ significance at $0.3\lesssim$ \dprojn\ $\lesssim 1$, followed by a reversal at the smallest pair separations. In contrast, previous statistical studies did not find signals of SFR suppression among dwarf galaxy pairs \citep{Stierwalt2015, Paudel2018}, possibly due to the lack of controlling for the \hi\ mass as we do. The trends do not change significantly if we divide the low-mass galaxies into primary and secondary ones (Figure \ref{fig:low-mass}a). 

To understand these trends, we show the correlations between \dsfr\ and \rgas\ for four subsamples divided by stellar mass and \dprojn\ (Figure \ref{fig:low-mass}b). For low-mass galaxies with \dprojn\ $>1$ (open blue circles), the SFR enhancement is around zero, similar to the trend for the parent sample. Once the dark matter halos encounter at \dprojn\ $\approx1$, the SFRs of low-mass galaxies become suppressed (filled blue circles). \dsfr\ then increases with decreasing \rgas, and finally exceeds zero when \rgas\ $<1$, i.e., when the \hi\ disks also encounter. The range of \rgas\ $< 1$ corresponds to \dprojn\ $\lesssim 0.2$ in Figure \ref{fig:low-mass}a, where \dsfr\ becomes positive. These results indicate that collisions of \hi\ disks strongly induce the SFR enhancement in low-mass galaxies. Compression of cold gas, coupled with gas inflows induced by shocks, might be responsible for this phenomenon \citep[e.g.,][]{Jog1992, Inoue2018, Moon2019}. The enhanced star formation in the interaction zone between the dwarf galaxies also fits this scenario \citep{Gao2023}. The increase of \dsfr\ with decreasing \rgas\ before \rgas\ $<1$ is possibly because the collision of hot gas halos \citep{Bekki2003}, which has similar though weaker effect as the collision of \hi\ disks. On the other hand, SFRs of low-mass galaxies in our sample are less affected by tidal interactions, as no significant correlation between \dsfr\ and $S_{\rm tid}$ is observed (Figure \ref{fig:high-mass}a).

After the halos meet but before the \hi\ disks collide (i.e., \dprojn\ $<1$ and \rgas\ $>1$), the SFRs of low-mass galaxies show a marginal suppression of $\sim 0.25$ dex, and are lower than those of the high-mass galaxies and the low-mass galaxies at \dprojn $>1$ with the same \rgas. Combining Figure \ref{fig:high_low_mass} and Figure \ref{fig:low-mass}a also indicates the underlying mechanisms for the suppression should prefer the low-mass galaxies over the high-mass ones, regardless of whether they are primary or secondary galaxies. However, due to limited sample size, we cannot further divide the sample by other parameters. A larger dataset is required to improve the statistics.\footnote{The greater scatter in scaling relations for low-mass galaxies may contribute to the large error bars. However, testing by increasing the number of controls for each galaxy indicates this effect to be minor.}

The SFR suppression was not reported in previous studies based on statistical samples, but is theoretically possible and supported by detailed studies of individual systems \citep{Kado-Fong2024}. Low-mass galaxies tend to be dark matter-dominated and less capable of restoring gas on a dynamically cold disk \citep{McConnachie2012, Pillepich2019}. Their \hi\ is more vulnerable to stellar feedback and external perturbations, even those from a secondary galaxy. These processes can re-distribute \hi\ into larger height or radius, temporarily reducing the amount of \hi\ linked to star formation \citep{Pearson2016,Pearson2018,Kado-Fong2024}. In addition, low-mass galaxies are less likely to harbor strong bars and spiral arms, making the mode-driven gas inflows inefficient \citep{Blumenthal2018} and the re-distributed \hi\ may only return well after the galaxies coalesce \citep{Pearson2018}. The limited spatial resolution of WALLABY prevents a detailed comparison with isolated dwarfs (but see \citet{Kim2023} for analysis of a smaller, better-resolved pair sample based on WALLABY data at redshift $z < 0.02$).

\subsection{Evolution of high-mass paired galaxies} \label{subsec:high-mass}

\begin{figure}
    \centering
    \includegraphics[width=0.46\textwidth]{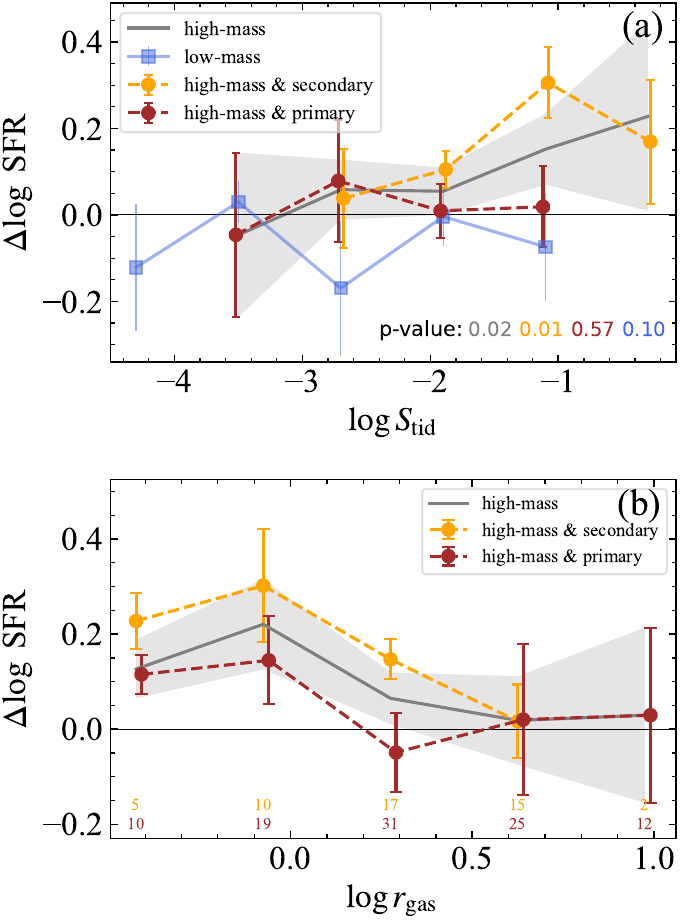}
    \caption{\textbf{Top}: SFR enhancement as a function of $\log S_{\rm tid}$. The gray line and blue lines represent high- and low-mass galaxies, respectively. The $p$-values of the correlations, corresponding to each sample, are listed at the bottom and share the same colors as the respective lines. \textbf{Bottom}: SFR enhancement as a function of \rgas\ for high-mass galaxies, similar to Figure \ref{fig:low-mass}a. Bins containing fewer than five galaxies are not shown.}
    \label{fig:high-mass}
\end{figure}

The evolution of high-mass paired galaxies is characterized by the continuous rise of SFR enhancement with decreasing pair separation, after their dark matter halos encounter (\dprojn\ $\approx1$; Figure \ref{fig:high_low_mass}). This scenario is qualitatively similar to the one described by \citet{Park2009}, who found that the H$\alpha$ equivalent width of late-type galaxy pairs rises as \dproj\ approaches the virial radius of their nearest neighbor. Together with the trend of increasing \dsfr\ with decreasing \dvn, it suggests that tidal interactions are playing a role here. This is more clearly demonstrated in Figure \ref{fig:high-mass}a: \dsfr\ of high-mass galaxies increases with the tidal parameter $S_{\rm tid}$ (gray line) while low-mass galaxies do not (blue line). 

The trend is stronger for secondary high-mass galaxies than for primary high-mass galaxies (Figure \ref{fig:high-mass}a). It implies that tidal perturbations significantly enhance the SFR in high-mass secondary galaxies, but cannot fully explain the weak enhancement of SFR in high-mass primary galaxies, even when the tidal strengths are at the same level. 
A difference in \dsfr\ is observed between primary and secondary galaxies for $\log r_{\rm gas} \lesssim 0.5$ (Figure \ref{fig:high-mass}b), well before the \hi\ disks overlap, suggesting that cold gas collision is not the primary trigger for the enhanced SFRs in secondary galaxies. Yet, the hydrodynamic effects of the hot gas halos can not be dismissed. Recent studies show that tidal interactions may trigger cold gas condensation from the hot gas halo around galaxies \citep{Sparre2022, Wang2023a}. In such a scenario, the diffuse gas condensed out of the hot gas halo tends to be captured by the more massive primary galaxy, but the newly accreted \hi\ is not capable of fueling star formation immediately, thus effectively leading to a suppressed \dsfr. This effect may cancel out with some of the tidally related SFR enhancing mechanisms, and lead to the lack of correlation between \dsfr\ and $S_{\rm tid}$ in primary massive galaxies.

The analysis above converges to a relatively important role of tidal effects for high-mass paired galaxies. This scenario can be connected with the anti-correlation between \dsfr\ and \hi-richness at $\Delta\log M_{\rm HI}\lesssim 0$ (bottom panel of Figure \ref{fig:high_low_mass}). The largely equilibrated states of low-redshift galaxies suggest that star formation fueling tends to be efficient in star-forming, \hi-rich, and massive galaxies \citep{Wang2020}, leaving little room for galaxy tidal interactions to enhance. This situation is opposite for the less \hi-rich galaxies. Moreover, tidally induced disk structures (e.g., bars) capable of driving gas inflows \citep{Barnes1991, Blumenthal2018} are more likely to form with lower gas fractions \citep{Masters2012}. Tidal effects may trigger disk instabilities in the cold gas and boost the star formation efficiency, which is otherwise much lower in a gas-poor dynamically hot disk when interactions are absent \citep{Martig2009}. From another perspective, the low \hi\ content may result from gas consumption by recently enhanced star formation and feedback processes. 

\begin{figure}
    \centering
    \includegraphics[width=0.47\textwidth]{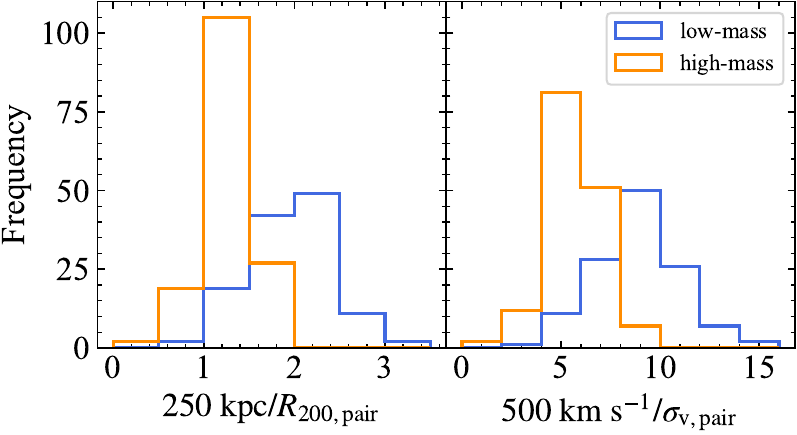}
    \caption{Distributions of the ratios $250~{\rm kpc}/R_{\rm 200, pair}$ (left) and $500~{\rm km~s^{-1}}/\sigma_{\rm v, pair}$ (right) for low-mass galaxies (blue) and high-mass galaxies (orange) in the parent sample.}
    \label{fig:R_sigma}
\end{figure}

Finally, we should be aware of the potential bias introduced by the maximum distances for our pair selection. Generally, a projected distance of 250 kpc corresponds to 1-2 $R_{\rm 200, pair}$ for high-mass galaxies (Figure \ref{fig:R_sigma}), making our analysis prefer low-mass galaxies at large normalized distances (\dprojn\ $\gtrsim 1.5$). We expect the bias to have minimal effects, given that no significant SFR enhancement is observed at this separation. The bias for the radial velocity offsets is even less pronounced, as $\sigma_{v,\rm pair}$ typically falls well below 500 $\rm km~s^{-1}$ (Figure \ref{fig:R_sigma}).

\subsection{Summary of the effects in gas-rich galaxy interactions}

\begin{figure*}
    \centering
    \includegraphics[width=\textwidth]{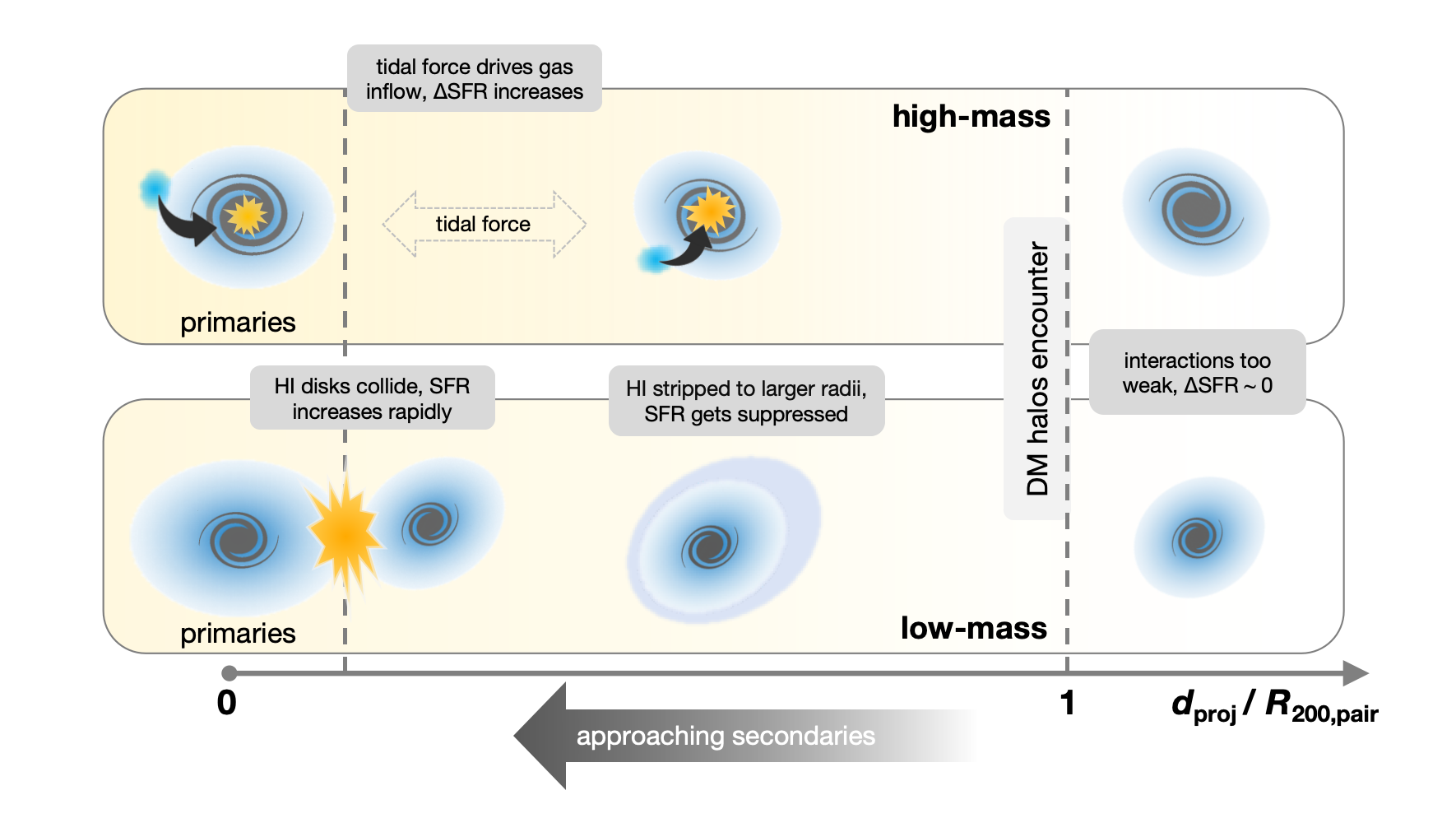}
    \caption{Schematic of the evolutionary pathways for gas-rich galaxy pairs. The gray spirals and the blue ellipses stand for the stellar and \hi\ disks, respectively. The explosion-shaped symbols represent the enhanced SFR, and the curved arrows represent possible gas inflows. Primary galaxies are fixed at \dprojn\ $=0$, while secondary galaxies approach the primaries from the right to the left. High- and low-mass galaxies are shown in the top and bottom rows, respectively.}
    \label{fig:pic}
\end{figure*}

Based on the previous discussion, we summarize the effects of tidal and hydrodynamic processes during the interactions of gas-rich galaxy pairs. We trace these effects from a projected distance of \dproj\ $> R_{\rm 200, pair}$ down to around $0.1R_{\rm 200, pair}$. The impact from larger-scale environments is minimized by selecting isolated pairs (Section \ref{subsubsec:environment}). A schematic plot of the interaction process is provided in Figure \ref{fig:pic}.

\textbf{(i) Tidal effects.} Tidal interactions affect the SFR enhancement at \dprojn\ $\lesssim 1$, where the dark matter halos of the galaxies have encountered. In high-mass galaxies with $\log M_\ast/M_\odot > 9$, \dsfr\ increases significantly with tidal strength. Tidal forces may act by inducing non-axisymmetric structures in the disks, especially at low \hi-richness, where the formation of such structures is facilitated. These structures, in turn, lead to gas inflows and enhance the central star formation. Conversely, low-mass galaxies exhibit a much weaker correlation between \dsfr\ and tidal strength, possibly due to their inability to hold the star-forming gas, and partly due to the disk structures in these systems.

\textbf{(ii) Hydrodynamic effects.} Strong interactions between the gas components are expected when the halos overlap, which agrees with the lack of a notable SFR enhancement (or suppression) at \dprojn\ $\gtrsim 1$. Hydrodynamic processes exhibit the strongest impact for low-mass galaxies at \rgas\ $\lesssim 1$, where collisions of the \hi\ disks greatly enhance the SFRs in these systems. At \rgas\ $> 1$, ram pressure exerted by the hot gas halo leads to gas compression, and collision between gas halos also leads to gas condensation and inflows, which may ultimately enhance the star formation. Cold gas collisions and ram pressure seem less important for high-mass galaxies.

It should be emphasized that the descriptions in Figure \ref{fig:pic} are highly simplified since we only focus on the dominant mechanisms in any specific stage. In reality, multiple mechanisms always act simultaneously during the interactions. Backsplashing paired galaxies evolve in the opposite direction as shown in Figure \ref{fig:pic}. They frequently show disturbed morphology and stripped \hi\ gas \citep[e.g.,][]{Yoon2017}, and are therefore less likely to be included in our sample, as also indicated by the small number of galaxies in Q4 (Figure \ref{fig:psd}).

\section{Conclusion} \label{sec:conclusion}
In this paper, we investigate the SFR enhancement of 278 \hi-rich galaxies in isolated galaxy pairs detected by WALLABY with redshift $z<0.075$ and stellar mass $7.6 < \log M_\ast/M_\odot < 11.2$. The pairs are selected to have projected distances \dproj\ $<250$ kpc and radial velocity offsets \dv\ $<500~\rm km~s^{-1}$. We develop an optical-based 3D deblending algorithm for \hi\ data cubes, which can effectively recover the \hi\ fluxes of close galaxy pairs when their 2D moment-0 maps are blended. For the first time, we trace the SFR enhancement of galaxy pairs across various interaction stages with their \hi\ fractions controlled, which allows us to further reveal the underlying physics during the interaction processes.

Our main conclusions are summarized as follows: 

\begin{enumerate}
    \item On average, the SFRs of gas-rich interacting galaxies in our sample are enhanced compared to control galaxies with similar stellar mass and \hi\ mass. The SFR enhancement increases with increasing pair proximity (Figure \ref{fig:sfr_depend}a, \ref{fig:sfr_depend}b). 
    \item The SFR enhancement depends weakly on stellar mass and stellar mass ratio for our parent sample. Galaxies with a major companion exhibit slightly higher \dsfr\ ($\approx 0.1$ dex) than those with minor companions (Figure \ref{fig:sfr_depend}f).
    \item Normalizing the separations of galaxy pairs with the halo properties ($R_{\rm 200, pair}$ and $\sigma_{v,\rm pair}$) enhances their correlations with \dsfr. \dsfr\ rises steeply as the normalized separations approach zero and becomes consistent with zero when the normalized separations exceed unity (Figure \ref{fig:sfr_depend}c, \ref{fig:sfr_depend}d). 
    \item High-mass ($M_\ast>10^9M_\odot$) and low-mass ($M_\ast<10^9M_\odot$) galaxies exhibit different behavior during the interactions. After the halos of the galaxy pairs encounter at \dprojn\ $\approx 1$, high-mass galaxies exhibit a continuous increase in the SFR enhancement, while low-mass galaxies show a marginal SFR suppression of $\sim 0.25$ dex at this stage at 1.4$\sigma$ significance. The SFRs of low-mass galaxies are only enhanced as their \hi\ disks overlap at \rgas\ $\lesssim1$ (Figure \ref{fig:low-mass}).
    \item High-mass galaxies are more sensitive to tidal perturbations than the low-mass ones, especially for the secondary galaxies of the pairs (Figure \ref{fig:high-mass}). High-mass galaxies also exhibit highly elevated SFRs ($\gtrsim 0.6$ dex) at low \hi-richness (Figure \ref{fig:high_low_mass}). Low-mass galaxies do not display such behavior as their high-mass counterparts.

\end{enumerate}

Based on these findings, we present a scenario outlining the evolution of SFR enhancement in \hi-rich galaxy pairs (Figure \ref{fig:pic}). In high-mass galaxies, tidal perturbations play more important roles in triggering star formation than in low-mass galaxies. Tidal force drives effective gas inflows into galactic centers at low \hi-richness and induces starbursts. For low-mass galaxies, tidal and ram pressure stripping move \hi\ into larger radii during the interactions, probably leading to a temporary SFR suppression after the halo encounters. As the \hi\ disks of the low-mass galaxies collide, gas compression and shocks induce intense star formation activities, resulting in an SFR enhancement level comparable to those in high-mass galaxies.

With the ongoing survey of WALLABY, much larger samples of gas-rich interacting systems will be available for analysis, and more galaxy properties can be explored with better statistics and greater dynamical ranges (e.g., galaxy morphology). In particular, the effects of galactic disk encounters are dependent on the specific orbit geometry involved \citep{Cox2008}, which may contribute to the scatter of \dsfr\ in this study.

The studies of star formation within interacting galaxies would also greatly benefit from synergies with CO observations \citep{Lee2022}, which provide more direct indicators of star-forming gas and was extensively used to study interacting galaxies in the past. These studies have hinted at the phase transition from \hi\ to $\rm H_2$ during the interactions as an important factor in affecting the SFRs \citep{Mirabel1989, Larson2016, Lisenfeld2019, Yu2024a}. Our deblending algorithm will greatly help to study multi-phase gas in merging systems.

\software{Astropy \citep{TheAstropyCollaboration2013,TheAstropyCollaboration2022},
            Photutils \citep{Bradley2022},
            scikit-image \citep{vanderWalt2014},
            SlicerAstro \citep{Punzo2017},
            SoFiA \citep{Serra2015, Westmeier2021}}

\section*{}
The authors thank the anonymous referee for useful comments that have improved this manuscript. We gratefully acknowledge Sara Ellison, Connor Bottrell, Juan Mardrid, Yago Ascasibar, and Thijs van der Hulst for helpful discussions or comments, and thank Chandrashekar Murugeshan for the \hi\ flux correction formula of WALLABY PDR2 data. JW thanks support of the research grants from National Science Foundation of China (NO. 12073002), Ministry of Science and Technology of the People's Republic of China (NO. 2022YFA1602902), and China Manned Space Project. SHOH acknowledges a support from the National Research Foundation of Korea (NRF) grant funded by the Korea government (Ministry of Science and ICT: MSIT) (No. RS-2022-00197685). XKC acknowledges the support from China Postdoctoral Science Foundation (2023M730094). KS acknowledges support from the Natural Sciences and Engineering Research Council of Canada (NSERC). Parts of this research were supported by the Australian Research Council Centre of Excellence for All Sky Astrophysics in 3 Dimensions (ASTRO 3D), through project number CE170100013. AB acknowledges support from the Centre National d'Etudes Spatiales (CNES), France.

This scientific work uses data obtained from Inyarrimanha Ilgari Bundara / the Murchison Radio-astronomy Observatory. We acknowledge the Wajarri Yamaji People as the Traditional Owners and native title holders of the Observatory site. CSIRO's ASKAP radio telescope is part of the Australia Telescope National Facility (\url{https://ror.org/05qajvd42}). Operation of ASKAP is funded by the Australian Government with support from the National Collaborative Research Infrastructure Strategy. ASKAP uses the resources of the Pawsey Supercomputing Research Centre. Establishment of ASKAP, Inyarrimanha Ilgari Bundara, the CSIRO Murchison Radio-astronomy Observatory and the Pawsey Supercomputing Research Centre are initiatives of the Australian Government, with support from the Government of Western Australia and the Science and Industry Endowment Fund.

WALLABY acknowledges technical support from the Australian SKA Regional Centre (AusSRC).

The Legacy Surveys consist of three individual and complementary projects: the Dark Energy Camera Legacy Survey (DECaLS; Proposal ID \#2014B-0404; PIs: David Schlegel and Arjun Dey), the Beijing-Arizona Sky Survey (BASS; NOAO Prop. ID \#2015A-0801; PIs: Zhou Xu and Xiaohui Fan), and the Mayall z-band Legacy Survey (MzLS; Prop. ID \#2016A-0453; PI: Arjun Dey). DECaLS, BASS and MzLS together include data obtained, respectively, at the Blanco telescope, Cerro Tololo Inter-American Observatory, NSF's NOIRLab; the Bok telescope, Steward Observatory, University of Arizona; and the Mayall telescope, Kitt Peak National Observatory, NOIRLab. Pipeline processing and analyses of the data were supported by NOIRLab and the Lawrence Berkeley National Laboratory (LBNL). The Legacy Surveys project is honored to be permitted to conduct astronomical research on Iolkam Du'ag (Kitt Peak), a mountain with particular significance to the Tohono O'odham Nation.

The Legacy Surveys imaging of the DESI footprint is supported by the Director, Office of Science, Office of High Energy Physics of the U.S. Department of Energy under Contract No. DE-AC02-05CH1123, by the National Energy Research Scientific Computing Center, a DOE Office of Science User Facility under the same contract; and by the U.S. National Science Foundation, Division of Astronomical Sciences under Contract No. AST-0950945 to NOAO.

\clearpage
\appendix

\section{Blended systems with extended \hi\ tails} \label{app:hi_tail}
\begin{figure*}
    \centering
    \includegraphics[width=\textwidth]{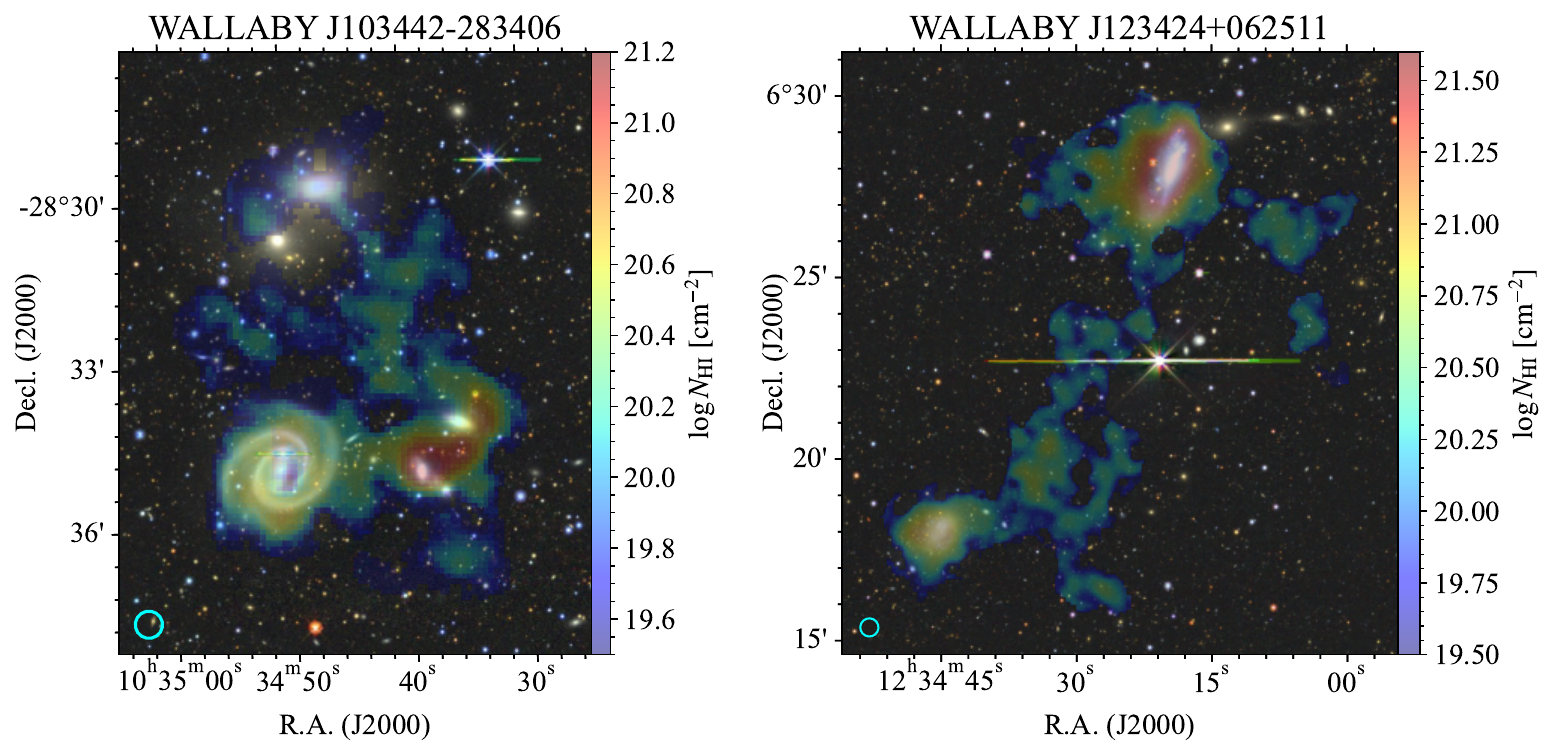}
    \caption{Optical images of WALLABY J103442-283406 (left) and WALLABY J123424+062511 (right) from the DESI Legacy Imaging Surveys, with the \hi\ moment-0 maps overlaid. Color-coding represents the \hi\ column density without correcting for \hi\ self-absorption. The cyan circles in the bottom left represent the synthesized beam of WALLABY.}
    \label{fig:hi_tail}
\end{figure*}

Figure \ref{fig:hi_tail} shows the two system with elongated \hi\ tails: WALLABY J103442-283406 and WALLABY J123424+062511. We can not attribute the extended \hi\ to either galaxy, and these two system are excluded from our analysis (Section \ref{subsec:wallaby}).

\section{Estimating the deblending error} \label{app:deblend}
We perform mock tests to quantify the uncertainties of the \hi\ fluxes of galaxies introduced by our deblending galaxies. This involves combining the \hi\ data cubes and optical images from two isolated galaxies in WALLABY to create simulated pairs of galaxies. In order to be better representative of the parent sample, we select simulated galaxy pairs that span an order of magnitude in both \hi\ flux ($\sim 9\times10^3$, $3\times10^4$, $8\times10^4\rm~Jy~Hz$) and \hi\ flux ratio ($\sim$ 1:1, 1:3, 1:10), resulting in nine pairs of galaxies in total. After employing the deblending algorithm on these mock data, we can estimate the deblending errors of \hi\ fluxes ($\Delta f_{\rm HI}$) by comparing the true fluxes to the fluxes recovered by the algorithm. For each simulated galaxy pair, we also alter the projected distance and the radial velocity offset to examine how the errors change with pair separation. The performance of the algorithm is quantified as the ratio between the deblending error $\Delta f_{\rm HI}$ and the \hi\ flux error due to observational noise ($\sigma_{f_{\rm HI}}$).

\begin{figure*}
    \centering
    \includegraphics[width=1\textwidth]{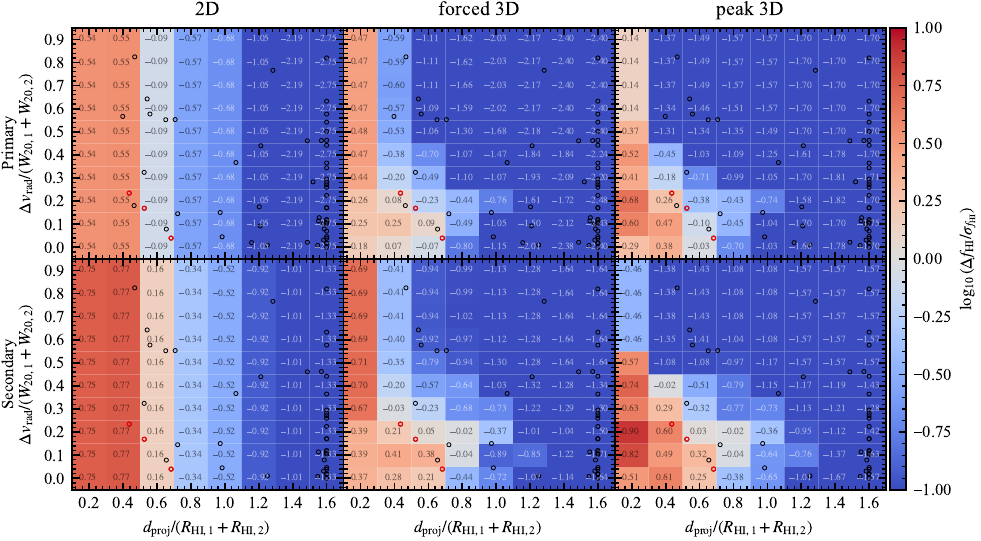}
    \caption{Errors of the deblending algorithm for galaxy pairs at different projected and radial velocity separations, normalized by the sum of their \hi\ disk sizes ($R_{\rm HI}$) and velocity widths of the \hi\ line profile measured at the 20\% level of the peaks ($W_{20}$), respectively. The panels share a consistent color coding that represents the mean errors of the nine simulated galaxy pairs, as also denoted by numerical values within the plot. The first row presents results for primary galaxies, and the second row is for secondary galaxies. The three columns, from left to right, represent the 2D, ``forced 3D'', and ``peak 3D'' methods, respectively. The small circles represent the position of galaxies in our parent sample, with black circles indicating galaxies deblended using the ``peak 3D'' method and red circles denoting those deblended with the ``forced 3D'' method. All the galaxy pairs with \dproj\ $/(R_{\rm HI,1}+R_{\rm HI,2})>1.6$ are placed at the abscissa of 1.6.}
    \label{fig:deblend_error}
\end{figure*}

Figure \ref{fig:deblend_error} summarizes the results of these mock tests. Overall, the ``peak 3D'' method generates the lowest errors among the three methods, except in the regions where the two galaxies are heavily blended and the ``forced 3D'' method performs slightly better. At large projected separations (\dproj\ $\gtrsim R_{\rm HI,1} + R_{\rm HI,2}$), the 2D approach is viable to recover the \hi\ flux of blended galaxies. But it performs notably worse than the 3D methods when dealing with close pairs, especially for pairs with large \dv. Primary galaxies show lower deblending errors than the secondary ones, mainly due to the higher $\sigma_{f_{\rm HI}}$ at the same $\Delta f_{\rm HI}$ for primary galaxies. 

The vast majority of the blended galaxies in our sample fall within the regions where the deblending error is substantially lower than the observational noise, as indicated by the small circles in Figure \ref{fig:deblend_error}. Figure \ref{fig:deblend_error} also validates our strategy of using the ``forced 3D'' method as an alternative when the ``peak 3D'' method fails, as the former produces smaller errors in the regions where %failures in ``peak 3D'' (red circles) begin to occur.

\section{Star-formation rate measurements} \label{app:control}

\begin{figure*}
    \centering
    \includegraphics[width=\textwidth]{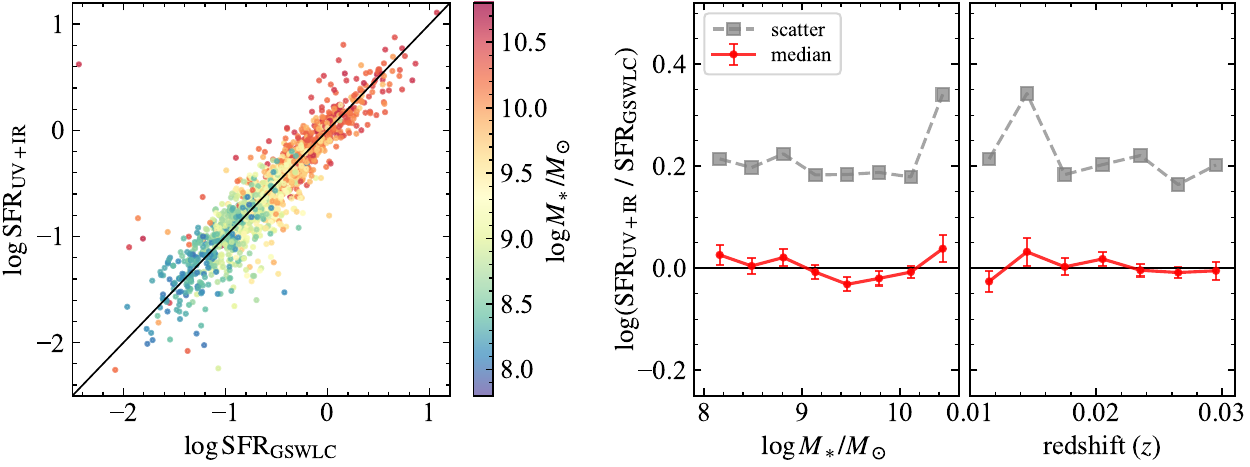}
    \caption{Consistency of the SFR measurements between the parent sample and the control sample. \textbf{Left}: Comparison of SFRs measured through our methods ($\rm SFR_{UV+IR}$) and SED fitting ($\rm SFR_{GSWLC}$) for the 1500 randomly selected galaxies (see text), color-coded by their stellar mass from the GSWLC-2 catalog \citep{Salim2018}. The black solid line along the diagonal represents the 1:1 relation. \textbf{Right}: Median deviation (red solid lines) and scatter (gray dashed lines) between the two SFR measurements as a function of stellar mass and redshift.}
    \label{figapp:sfr}
\end{figure*}

To examine the consistency between SFRs of the parent and the control samples, we select 1500 galaxies randomly from the control pool with redshift $0.01<z<0.03$ and a uniform stellar mass distribution covering $8.0<\log M_\ast/M_\odot<10.6$. These galaxies are chosen to be representative of the parent sample and to ensure statistical significance. We measure the SFRs of these galaxies using the UV+IR luminosity, following the identical procedures described in Section \ref{subsubsec:sfr}. The comparison of the SFRs obtained from our measurements and from \citet{Salim2018} is shown in Figure \ref{figapp:sfr}. We see that the systematic difference of the SFRs is consistent with zero, and the scatter roughly remains constant around 0.2 dex. The consistency validates our determination of the SFR enhancement by subtracting SFRs obtained with the two methods (Equation \ref{eq:dsfr}). Our main conclusions are robust despite the different methods in SFR measurements.

\section{Effects of environment cut on the control sample} \label{app:env}

\begin{figure*}
    \centering
    \includegraphics[width=0.55\linewidth]{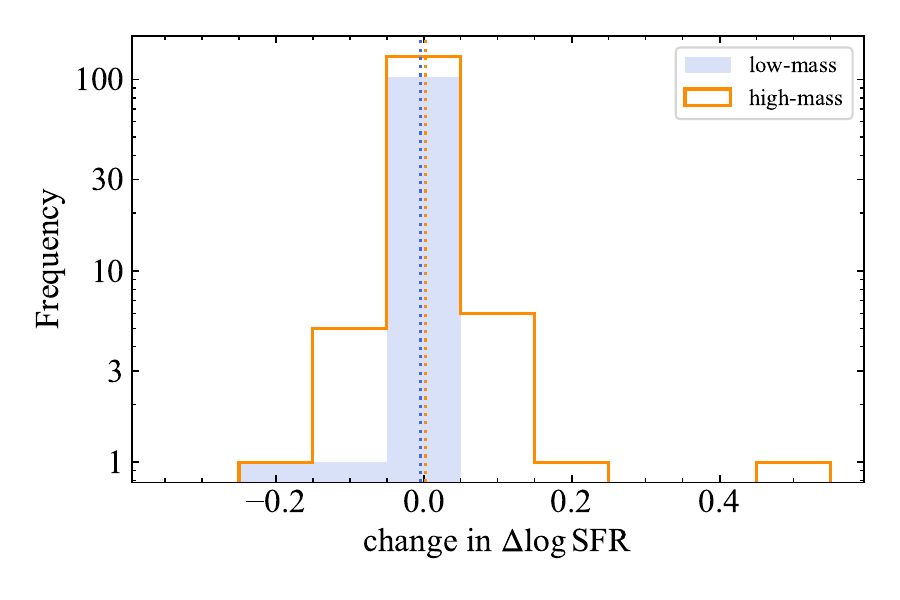}
    \caption{Histograms of the change in \dsfr\ after applying the environment cut to the control sample. The vertical lines represent the mean value of the high/low mass subsets.}
    \label{fig:env}
\end{figure*}

In Section \ref{subsec:control}, we noted that our control pool includes members of massive galaxy groups, which may raise concerns on whether these group members, often characterized by suppressed SFRs, could artificially boost the \dsfr\ of our parent sample.

\begin{figure*}
    \centering
    \includegraphics[width=\linewidth]{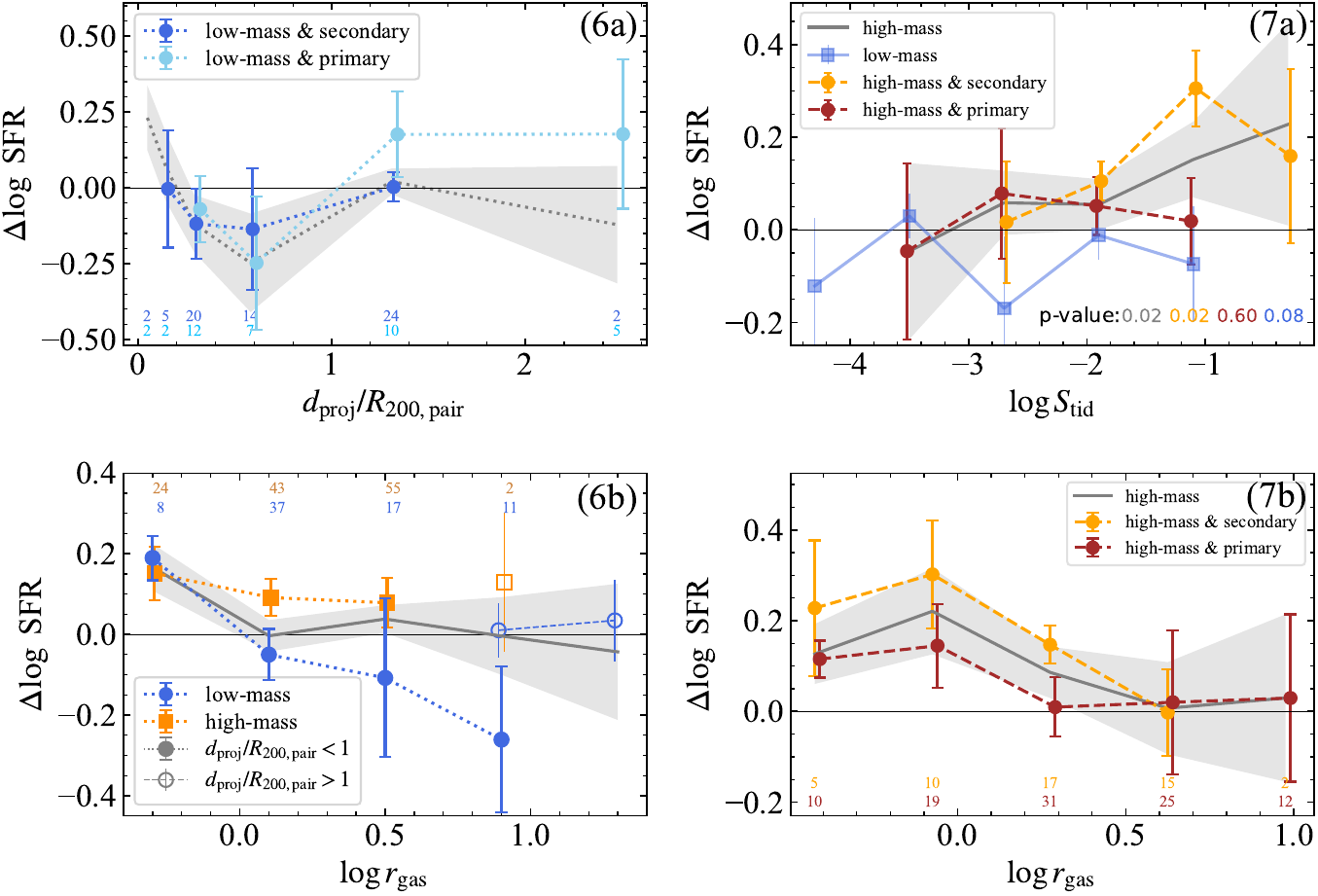}
    \caption{Replicates of Figures \ref{fig:low-mass} and \ref{fig:high-mass}, except that galaxies in massive groups are removed from the control pool. No significant offset is observed compared to Figures \ref{fig:low-mass} and \ref{fig:high-mass}.}
    \label{fig:67}
\end{figure*}

Figures \ref{fig:high_low_mass} to \ref{fig:high-mass} illustrate that galaxy pairs with the weakest interactions exhibit neither SFR enhancement nor suppression, which validates our control process. To further confirm this, we apply the same isolation criteria described in Section \ref{subsubsec:environment} to the control pool and compare the updated \dsfr\ values with the original ones. We find that only 5\% of the galaxies are affected, both in the control pool (676 out of 14449) and the parent sample (14 out of 278; Figure \ref{fig:env}). There is also no systematic offset in \dsfr\ for the 14 paired galaxies, and all the results presented in this paper remain valid. As an example, Figure \ref{fig:67} nicely reproduces the results shown in Figures \ref{fig:low-mass} and \ref{fig:high-mass} with the updated control pool.

\section{Catalog of Paired Galaxies} \label{app:catalog}

Basic properties of individual galaxies in the parent sample are provided in Table \ref{tab:pair}.

%\begin{longrotatetable}
\begin{deluxetable*}{ccrcrrcccrrrrc}
\tablecaption{Properties of individual paired galaxies in the parent sample. \label{tab:pair}}
\tablewidth{\textwidth}
\decimalcolnumbers
\tabletypesize{\scriptsize}
\tablehead{name & R.A. & Dec. & $z$ & $\log M_\ast$ & log SFR & GALEX & $\log M_{\rm HI}$ & deblend & $g_{\rm petro}$ & $r_{\rm petro}$ & $i_{\rm petro}$ & $R_{\rm 25,g}$ & control \\
 & (deg) & (deg) &  & ($M_\odot$) & ($M_\odot\rm~yr^{-1}$) & flag & ($M_\odot$) & flag & (mag) & (mag) & (mag) & (arcsec) & flag }
\startdata
WALLABY\_J101443-263328 & 153.6646 & -26.5625 & 0.05393 &  9.61 &  0.14 & F & 10.01 & 1 & 16.76 & 16.32 & 16.20 &   9.1 & 1 \\
WALLABY\_J101443-263328 & 153.6884 & -26.5551 & 0.05363 &  9.78 & -0.02 & F & 10.23 & 1 & 17.59 & 16.93 & 16.70 &  12.6 & 1 \\
WALLABY\_J101448-274240 & 153.6983 & -27.7117 & 0.01410 &  8.30 & -1.17 & F &  9.34 & 1 & 16.94 & 16.57 & 16.46 &  13.4 & 1 \\
WALLABY\_J101448-274240 & 153.7257 & -27.7082 & 0.01420 &  7.73 & -0.74 & N &  8.88 & 1 & 17.29 & 17.11 & 17.11 &   6.2 & 1 \\
WALLABY\_J101834-281550 & 154.6438 & -28.2683 & 0.04080 &  9.12 & -0.12 & F &  9.71 & 1 & 17.10 & 16.72 & 16.62 &   9.0 & 1 \\
WALLABY\_J101834-281550 & 154.6490 & -28.2646 & 0.04077 &  8.79 & -0.32 & F &  9.61 & 1 & 17.11 & 16.87 & 16.83 &  11.3 & 1 \\
WALLABY\_J101945-272719 & 154.9413 & -27.4582 & 0.03626 &  9.94 & -0.05 & F &  9.68 & 1 & 16.11 & 15.49 & 15.29 &  12.5 & 1 \\
WALLABY\_J101945-272719 & 154.9391 & -27.4517 & 0.03672 &  9.50 & -0.05 & F &  9.58 & 1 & 16.53 & 16.04 & 15.90 &  12.2 & 1 \\
WALLABY\_J102019-285220 & 155.0796 & -28.8738 & 0.03133 & 10.67 & -0.16 & N &  9.67 & 1 & 15.11 & 14.30 & 13.98 &  49.5 & 1 \\
WALLABY\_J102019-285220 & 155.0931 & -28.8705 & 0.03074 & 10.15 &  0.36 & F &  9.49 & 1 & 15.39 & 14.74 & 14.53 &  20.2 & 1 \\
WALLABY\_J102054-263844 & 155.2233 & -26.6460 & 0.04121 &  8.90 & -0.17 & F &  9.81 & 1 & 17.20 & 16.74 & 16.84 &  10.5 & 1 \\
WALLABY\_J102054-263844 & 155.2394 & -26.6431 & 0.04121 &  8.28 & -1.03 & F &  9.58 & 1 & 18.66 & 18.53 & 18.32 &   9.1 & 1 \\
WALLABY\_J103540-284607 & 158.9026 & -28.7686 & 0.03047 &  8.93 & -0.30 & F &  9.88 & 1 & 16.76 & 16.46 & 16.34 &  16.5 & 1 \\
WALLABY\_J103540-284607 & 158.9368 & -28.7691 & 0.03065 &  9.89 & -0.42 & F &  9.81 & 1 & 15.99 & 15.36 & 15.14 &  17.1 & 1 \\
WALLABY\_J104513-262755 & 161.3050 & -26.4614 & 0.01432 &  7.77 & -1.66 & F &  9.04 & 1 & 17.93 & 17.53 & 17.55 &   8.5 & 1
\enddata
\tablecomments{Columns: (1) Source name from WALLABY. (2)(3) Right ascension and declination from optical images. (4) Redshift from WALLABY \hi\ data. (5) Stellar mass derived from luminosity and mass-to-light ratio (Section \ref{subsubsec:mass}) (6) Star formation rate derived from UV+IR luminosity (Section \ref{subsubsec:sfr}). (7) Note on GALEX data used for measuring the SFR. ``F'': FUV, ``N'': NUV, ``nd'': non-detection in both bands, ``--'': not in GALEX footprints. (8) \hi\ mass from WALLABY. (9) Blending status of the \hi\ detection: 0 = no blending, 1 = deblending is performed to obtain the \hi\ mass (Section \ref{sec:deblend}). (10)--(12) $g$, $r$, and $i$ bands Petrosian magnitudes (Section \ref{subsubsec:mass}). (13) Semi-major axis measured at 25 $\rm mag~arcsec^{-2}$ in $g$ band. (14) Status of control. 1 = with five control galaxies, 0 = fewer than five control galaxies. Null values are filled with -99.00.
\\(This table is available in its entirety in machine-readable form.)}
\end{deluxetable*}
%\end{longrotatetable}

\section{Visualization of the 3D Deblending} \label{app:3D}

\begin{figure*}
    \centering
    \includegraphics[width=0.7\textwidth]{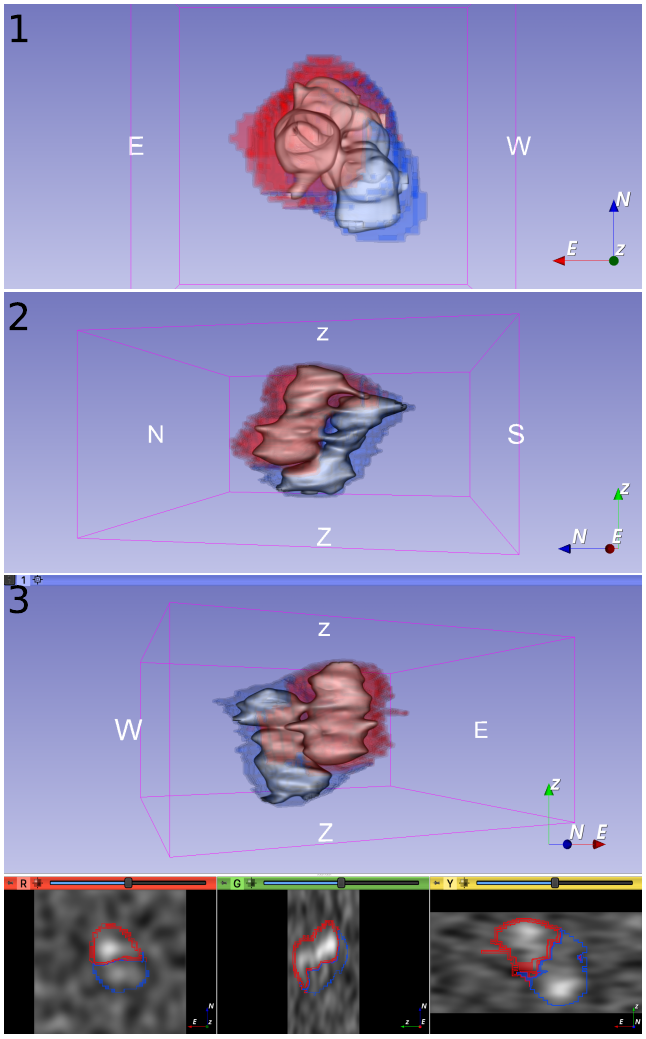}
    \caption{Snapshots of 3D visualization of the deblended \hi\ data cube (Figure \ref{fig:deblend}h) using \texttt{SlicerAstro} \citep{Punzo2017}.  The three larger panels display the \hi\ distribution and deblending results in different projections. The red and blue surfaces depict the 3D envelopes of the segments corresponding to the two galaxies within the \texttt{SoFiA} mask \citep{Westmeier2022}, and the gray surface represent a constant \hi\ flux density. The three smaller plots at the bottom exhibit three orthogonal slices of the data cube. The viewing angles are indicated by colored arrows in the lower right corner of each panel. An animation of this figure is available online at \url{https://github.com/BetaGem/wallaby-galaxy-pair/blob/main/Figure13.mp4}, showing the continuous rotation of the data cube in $360^\circ$ along the frequency axis.}
    \label{fig:3D}
\end{figure*}

Figure \ref{fig:3D} provides a three-dimensional visualization of the deblending results as shown in Figure \ref{fig:deblend}h. The visualization is produced by the software \texttt{SlicerAstro} \citep{Punzo2017}. The \hi\ content of the two galaxies can be clearly separated in panel 2, but is blended in the moment-0 map (panel 1).

\clearpage
\bibliography{Huang+24}{}

\begin{thebibliography}{}
\expandafter\ifx\csname natexlab\endcsname\relax\def\natexlab#1{#1}\fi
\providecommand{\url}[1]{\href{#1}{#1}}
\providecommand{\dodoi}[1]{doi:~\href{http://doi.org/#1}{\nolinkurl{#1}}}
\providecommand{\doeprint}[1]{\href{http://ascl.net/#1}{\nolinkurl{http://ascl.net/#1}}}
\providecommand{\doarXiv}[1]{\href{https://arxiv.org/abs/#1}{\nolinkurl{https://arxiv.org/abs/#1}}}

\bibitem[{Arp(1966)}]{Arp1966}
Arp, H. 1966, ApJS, 14, 1, \dodoi{10.1086/190147}

\bibitem[{Bacchini {et~al.}(2019)Bacchini, Fraternali, Iorio, \&
  Pezzulli}]{Bacchini2019}
Bacchini, C., Fraternali, F., Iorio, G., \& Pezzulli, G. 2019, A\&A, 622, A64,
  \dodoi{10.1051/0004-6361/201834382}

\bibitem[{Barnes(1992)}]{Barnes1992}
Barnes, J.~E. 1992, ApJ, 393, 484, \dodoi{10.1086/171522}

\bibitem[{Barnes \& Hernquist(1991)}]{Barnes1991}
Barnes, J.~E., \& Hernquist, L.~E. 1991, ApJ, 370, L65, \dodoi{10.1086/185978}

\bibitem[{Behroozi {et~al.}(2019)Behroozi, Wechsler, Hearin, \&
  Conroy}]{Behroozi2019}
Behroozi, P., Wechsler, R.~H., Hearin, A.~P., \& Conroy, C. 2019, \mnras, 488,
  3143, \dodoi{10.1093/mnras/stz1182}

\bibitem[{Bekki(2008)}]{Bekki2008}
Bekki, K. 2008, \mnras, 388, L10, \dodoi{10.1111/j.1745-3933.2008.00489.x}

\bibitem[{Bekki \& Couch(2003)}]{Bekki2003}
Bekki, K., \& Couch, W.~J. 2003, Astrophys. J., 596, L13,
  \dodoi{10.1086/379054}

\bibitem[{Bertin \& Arnouts(1996)}]{Bertin1996}
Bertin, E., \& Arnouts, S. 1996, \aaps, 117, 393, \dodoi{10.1051/aas:1996164}

\bibitem[{Bianchi {et~al.}(2017)Bianchi, Shiao, \& Thilker}]{Bianchi2017}
Bianchi, L., Shiao, B., \& Thilker, D. 2017, \apjs, 230, 24,
  \dodoi{10.3847/1538-4365/aa7053}

\bibitem[{Bigiel {et~al.}(2010)Bigiel, Leroy, Walter, Blitz, Brinks, {de Blok},
  \& Madore}]{Bigiel2010}
Bigiel, F., Leroy, A., Walter, F., {et~al.} 2010, \aj, 140, 1194,
  \dodoi{10.1088/0004-6256/140/5/1194}

\bibitem[{Blanton {et~al.}(2011)Blanton, Kazin, Muna, Weaver, \&
  {Price-Whelan}}]{Blanton2011}
Blanton, M.~R., Kazin, E., Muna, D., Weaver, B.~A., \& {Price-Whelan}, A. 2011,
  \aj, 142, 31, \dodoi{10.1088/0004-6256/142/1/31}

\bibitem[{Blumenthal \& Barnes(2018)}]{Blumenthal2018}
Blumenthal, K.~A., \& Barnes, J.~E. 2018, \mnras, 479, 3952,
  \dodoi{10.1093/mnras/sty1605}

\bibitem[{Bottrell {et~al.}(2023)Bottrell, Yesuf, Popping, Omori, Tang, Ding,
  Pillepich, Nelson, Eisert, Gao, Goulding, Kalita, Luo, Greene, Shi, \&
  Silverman}]{Bottrell2023}
Bottrell, C., Yesuf, H.~M., Popping, G., {et~al.} 2023, \mnras,
  \dodoi{10.1093/mnras/stad2971}

\bibitem[{Bradley {et~al.}(2022)Bradley, Sip{\H{o}}cz, Robitaille, Tollerud,
  Vin{\'i}cius, Deil, Barbary, Wilson, Busko, Donath, G{\"u}nther, Cara, Lim,
  Me{\ss}linger, Conseil, Bostroem, Droettboom, Bray, Andersen~Bratholm,
  Barentsen, Craig, Rathi, Pascual, Perren, Georgiev, {De Val-Borro},
  Kerzendorf, Bach, Quint, \& Souchereau}]{Bradley2022}
Bradley, L., Sip{\H{o}}cz, B., Robitaille, T., {et~al.} 2022,
  Astropy/{{Photutils}}: 1.5.0,  Zenodo, \dodoi{10.5281/zenodo.6825092}

\bibitem[{Brown {et~al.}(2014)Brown, Jarrett, \& Cluver}]{Brown2014}
Brown, M. J.~I., Jarrett, T.~H., \& Cluver, M.~E. 2014, \pasa, 31, e049,
  \dodoi{10.1017/pasa.2014.44}

\bibitem[{Brown {et~al.}(2023)Brown, Patton, Ellison, \& Faria}]{Brown2023a}
Brown, W., Patton, D.~R., Ellison, S.~L., \& Faria, L. 2023, \mnras, 522, 5107,
  \dodoi{10.1093/mnras/stad1314}

\bibitem[{Bustamante {et~al.}(2020)Bustamante, Ellison, Patton, \&
  Sparre}]{Bustamante2020}
Bustamante, S., Ellison, S.~L., Patton, D.~R., \& Sparre, M. 2020, \mnras, 494,
  3469, \dodoi{10.1093/mnras/staa1025}

\bibitem[{Cao {et~al.}(2016)Cao, Xu, Domingue, Buat, Cheng, Gao, Huang,
  Jarrett, Lisenfeld, Lu, Mazzarella, Sun, Wu, Yun, Ronca, \&
  Jacques}]{Cao2016}
Cao, C., Xu, C.~K., Domingue, D., {et~al.} 2016, ApJS, 222, 16,
  \dodoi{10.3847/0067-0049/222/2/16}

\bibitem[{Cerdosino {et~al.}(2024)Cerdosino, O'Mill, Rodriguez, Taverna,
  Sodr{\'e}~Jr, Telles, {M{\'e}ndez-Hern{\'a}ndez}, Schoenell, Ribeiro, Kanaan,
  \& {de Oliveira}}]{Cerdosino2024}
Cerdosino, M.~C., O'Mill, A.~L., Rodriguez, F., {et~al.} 2024, \mnras, 528,
  4993, \dodoi{10.1093/mnras/stae334}

\bibitem[{Chabrier(2003)}]{Chabrier2003}
Chabrier, G. 2003, PUBL ASTRON SOC PAC, 115, 763, \dodoi{10.1086/376392}

\bibitem[{Chilingarian {et~al.}(2010)Chilingarian, Melchior, \&
  Zolotukhin}]{Chilingarian2010}
Chilingarian, I.~V., Melchior, A.-L., \& Zolotukhin, I.~Y. 2010, \mnras, no,
  \dodoi{10.1111/j.1365-2966.2010.16506.x}

\bibitem[{Cimatti {et~al.}(2020)Cimatti, Fraternali, \& Nipoti}]{Cimatti2020}
Cimatti, A., Fraternali, F., \& Nipoti, C. 2020, Introduction to {{Galaxy
  Formation}} and {{Evolution}}: {{From Primordial Gas}} to {{Present-Day
  Galaxies}}

\bibitem[{Condon {et~al.}(1998)Condon, Cotton, Greisen, Yin, Perley, Taylor, \&
  Broderick}]{Condon1998}
Condon, J.~J., Cotton, W.~D., Greisen, E.~W., {et~al.} 1998, Astron. J., 115,
  1693, \dodoi{10.1086/300337}

\bibitem[{Conselice(2003)}]{Conselice2003}
Conselice, C.~J. 2003, \apjs, 147, 1, \dodoi{10.1086/375001}

\bibitem[{Cortese {et~al.}(2021)Cortese, Catinella, \& Smith}]{Cortese2021}
Cortese, L., Catinella, B., \& Smith, R. 2021, \pasa, 38, e035,
  \dodoi{10.1017/pasa.2021.18}

\bibitem[{Cox {et~al.}(2008)Cox, Jonsson, Somerville, Primack, \&
  Dekel}]{Cox2008}
Cox, T.~J., Jonsson, P., Somerville, R.~S., Primack, J.~R., \& Dekel, A. 2008,
  \mnras, 384, 386, \dodoi{10.1111/j.1365-2966.2007.12730.x}

\bibitem[{Davies {et~al.}(2015)Davies, Robotham, Driver, Alpaslan, Baldry,
  {Bland-Hawthorn}, Brough, Brown, Cluver, Drinkwater, Foster, Grootes,
  Konstantopoulos, {Lara-L{\'o}pez}, {L{\'o}pez-S{\'a}nchez}, Loveday, Meyer,
  Moffett, Norberg, Owers, Popescu, De~Propris, Sharp, Tuffs, Wang, Wilkins,
  Dunne, Bourne, \& Smith}]{Davies2015}
Davies, L. J.~M., Robotham, A. S.~G., Driver, S.~P., {et~al.} 2015, \mnras,
  452, 616, \dodoi{10.1093/mnras/stv1241}

\bibitem[{Dey {et~al.}(2019)Dey, Schlegel, Lang, Blum, Burleigh, Fan, Findlay,
  Finkbeiner, Herrera, Juneau, Landriau, Levi, McGreer, Meisner, Myers,
  Moustakas, Nugent, Patej, Schlafly, Walker, Valdes, Weaver, Y{\`e}che, Zou,
  Zhou, Abareshi, Abbott, Abolfathi, Aguilera, Alam, Allen, Alvarez, Annis,
  Ansarinejad, Aubert, Beechert, Bell, BenZvi, Beutler, Bielby, Bolton,
  Brice{\~n}o, {Buckley-Geer}, Butler, Calamida, Carlberg, Carter, Casas,
  Castander, Choi, Comparat, Cukanovaite, Delubac, DeVries, Dey, Dhungana,
  Dickinson, Ding, Donaldson, Duan, Duckworth, Eftekharzadeh, Eisenstein,
  Etourneau, Fagrelius, Farihi, Fitzpatrick, {Font-Ribera}, Fulmer,
  G{\"a}nsicke, Gaztanaga, George, Gerdes, Gontcho, Gorgoni, Green, Guy,
  Harmer, Hernandez, Honscheid, Huang, James, Jannuzi, Jiang, Joyce, Karcher,
  Karkar, Kehoe, Kneib, {Kueter-Young}, Lan, Lauer, Le~Guillou, Le~Van~Suu,
  Lee, Lesser, Perreault~Levasseur, Li, Mann, Marshall,
  {Mart{\'i}nez-V{\'a}zquez}, Martini, {du Mas des Bourboux}, McManus, Meier,
  M{\'e}nard, Metcalfe, {Mu{\~n}oz-Guti{\'e}rrez}, Najita, Napier, Narayan,
  Newman, Nie, Nord, Norman, Olsen, Paat, {Palanque-Delabrouille}, Peng,
  Poppett, Poremba, Prakash, Rabinowitz, Raichoor, Rezaie, Robertson, Roe,
  Ross, Ross, Rudnick, Safonova, Saha, S{\'a}nchez, Savary, Schweiker, Scott,
  Seo, Shan, Silva, Slepian, Soto, Sprayberry, Staten, Stillman, Stupak,
  Summers, Sien~Tie, Tirado, {Vargas-Maga{\~n}a}, Vivas, Wechsler, Williams,
  Yang, Yang, Yapici, Zaritsky, Zenteno, Zhang, Zhang, Zhou, \& Zhou}]{Dey2019}
Dey, A., Schlegel, D.~J., Lang, D., {et~al.} 2019, \aj, 157, 168,
  \dodoi{10.3847/1538-3881/ab089d}

\bibitem[{Dickey \& Lockman(1990)}]{Dickey1990}
Dickey, J.~M., \& Lockman, F.~J. 1990, \araa, 28, 215,
  \dodoi{10.1146/annurev.aa.28.090190.001243}

\bibitem[{Ellison {et~al.}(2018)Ellison, Catinella, \& Cortese}]{Ellison2018}
Ellison, S.~L., Catinella, B., \& Cortese, L. 2018, \mnras, 478, 3447,
  \dodoi{10.1093/mnras/sty1247}

\bibitem[{Ellison {et~al.}(2008)Ellison, Patton, Simard, \&
  McConnachie}]{Ellison2008}
Ellison, S.~L., Patton, D.~R., Simard, L., \& McConnachie, A.~W. 2008, \aj,
  135, 1877, \dodoi{10.1088/0004-6256/135/5/1877}

\bibitem[{Ellison {et~al.}(2019)Ellison, Viswanathan, Patton, Bottrell,
  McConnachie, Gwyn, \& Cuillandre}]{Ellison2019}
Ellison, S.~L., Viswanathan, A., Patton, D.~R., {et~al.} 2019, \mnras, 487,
  2491, \dodoi{10.1093/mnras/stz1431}

\bibitem[{Ellison {et~al.}(2022)Ellison, Wilkinson, Woo, Leung, Wild, Bickley,
  Patton, Quai, \& Gwyn}]{Ellison2022}
Ellison, S.~L., Wilkinson, S., Woo, J., {et~al.} 2022, \mnras, 517, L92,
  \dodoi{10.1093/mnrasl/slac109}

\bibitem[{Engler {et~al.}(2023)Engler, Pillepich, Joshi, Pasquali, Nelson, \&
  Grebel}]{Engler2023}
Engler, C., Pillepich, A., Joshi, G.~D., {et~al.} 2023, \mnras, 522, 5946,
  \dodoi{10.1093/mnras/stad1357}

\bibitem[{Evrard {et~al.}(2008)Evrard, Bialek, Busha, White, Habib, Heitmann,
  Warren, Rasia, Tormen, Moscardini, Power, Jenkins, Gao, Frenk, Springel,
  White, \& Diemand}]{Evrard2008}
Evrard, A.~E., Bialek, J., Busha, M., {et~al.} 2008, ApJ, 672, 122,
  \dodoi{10.1086/521616}

\bibitem[{Feng {et~al.}(2020)Feng, Shen, Yuan, Riffel, \& Pan}]{Feng2020}
Feng, S., Shen, S.-Y., Yuan, F.-T., Riffel, R.~A., \& Pan, K. 2020, \apj, 892,
  L20, \dodoi{10.3847/2041-8213/ab7dba}

\bibitem[{Finkbeiner {et~al.}(2016)Finkbeiner, Schlafly, Schlegel, Padmanabhan,
  Juri{\'c}, Burgett, Chambers, Denneau, Draper, Flewelling, Hodapp, Kaiser,
  Magnier, Metcalfe, Morgan, Price, Stubbs, \& Tonry}]{Finkbeiner2016}
Finkbeiner, D.~P., Schlafly, E.~F., Schlegel, D.~J., {et~al.} 2016, ApJ, 822,
  66, \dodoi{10.3847/0004-637X/822/2/66}

\bibitem[{For {et~al.}(2023)For, Spekkens, {Staveley-Smith}, Bekki,
  Karunakaran, Catinella, Koribalski, {Lee-Waddell}, Madrid, Murugeshan, Rhee,
  Westmeier, Wong, Zaritsky, \& Donnerstein}]{For2023}
For, B.~Q., Spekkens, K., {Staveley-Smith}, L., {et~al.} 2023, \mnras, 526,
  3130, \dodoi{10.1093/mnras/stad2921}

\bibitem[{{Gaia Collaboration} {et~al.}(2023){Gaia Collaboration}, Vallenari,
  Brown, Prusti, {de Bruijne}, Arenou, Babusiaux, Biermann, Creevey, Ducourant,
  Evans, Eyer, Guerra, Hutton, Jordi, Klioner, Lammers, Lindegren, Luri,
  Mignard, Panem, Pourbaix, Randich, Sartoretti, Soubiran, Tanga, Walton,
  {Bailer-Jones}, Bastian, Drimmel, Jansen, Katz, Lattanzi, {van Leeuwen},
  Bakker, Cacciari, Casta{\~n}eda, De~Angeli, Fabricius, Fouesneau, Fr{\'e}mat,
  Galluccio, Guerrier, Heiter, Masana, Messineo, Mowlavi, Nicolas,
  Nienartowicz, Pailler, Panuzzo, Riclet, Roux, Seabroke, Sordo, Th{\'e}venin,
  {Gracia-Abril}, Portell, Teyssier, Altmann, Andrae, Audard, {Bellas-Velidis},
  Benson, Berthier, Blomme, Burgess, Busonero, Busso, C{\'a}novas, Carry,
  Cellino, Cheek, Clementini, Damerdji, Davidson, {de Teodoro},
  Nu{\~n}ez~Campos, Delchambre, Dell'Oro, Esquej,
  {Fern{\'a}ndez-Hern{\'a}ndez}, Fraile, Garabato, {Garc{\'i}a-Lario}, Gosset,
  Haigron, Halbwachs, Hambly, Harrison, Hern{\'a}ndez, Hestroffer, Hodgkin,
  Holl, Jan{\ss}en, {Jevardat de Fombelle}, Jordan, {Krone-Martins}, Lanzafame,
  L{\"o}ffler, Marchal, Marrese, Moitinho, Muinonen, Osborne, Pancino, Pauwels,
  {Recio-Blanco}, Reyl{\'e}, Riello, Rimoldini, Roegiers, Rybizki, Sarro,
  Siopis, Smith, Sozzetti, Utrilla, {van Leeuwen}, Abbas, {\'A}brah{\'a}m,
  Abreu~Aramburu, Aerts, Aguado, Ajaj, {Aldea-Montero}, Altavilla, {\'A}lvarez,
  Alves, Anders, Anderson, Anglada~Varela, Antoja, Baines, Baker,
  {Balaguer-N{\'u}{\~n}ez}, Balbinot, Balog, Barache, Barbato, Barros, Barstow,
  Bartolom{\'e}, Bassilana, Bauchet, Becciani, Bellazzini, Berihuete, Bernet,
  Bertone, Bianchi, Binnenfeld, {Blanco-Cuaresma}, Blazere, Boch, Bombrun,
  Bossini, Bouquillon, Bragaglia, Bramante, Breedt, Bressan, Brouillet,
  Brugaletta, Bucciarelli, Burlacu, Butkevich, Buzzi, Caffau, Cancelliere,
  {Cantat-Gaudin}, Carballo, Carlucci, Carnerero, Carrasco, Casamiquela,
  Castellani, {Castro-Ginard}, Chaoul, Charlot, Chemin, Chiaramida, Chiavassa,
  Chornay, Comoretto, Contursi, Cooper, Cornez, Cowell, Crifo, Cropper, Crosta,
  Crowley, Dafonte, Dapergolas, David, David, {de Laverny}, De~Luise, De~March,
  De~Ridder, {de Souza}, {de Torres}, {del Peloso}, {del Pozo}, Delbo, Delgado,
  Delisle, Demouchy, Dharmawardena, Di~Matteo, Diakite, Diener, Distefano,
  Dolding, Edvardsson, Enke, Fabre, Fabrizio, Faigler, Fedorets, Fernique,
  Fienga, Figueras, Fournier, Fouron, Fragkoudi, Gai, {Garcia-Gutierrez},
  {Garcia-Reinaldos}, {Garc{\'i}a-Torres}, Garofalo, Gavel, Gavras, Gerlach,
  Geyer, Giacobbe, Gilmore, Girona, Giuffrida, Gomel, Gomez,
  {Gonz{\'a}lez-N{\'u}{\~n}ez}, {Gonz{\'a}lez-Santamar{\'i}a},
  {Gonz{\'a}lez-Vidal}, Granvik, Guillout, Guiraud,
  {Guti{\'e}rrez-S{\'a}nchez}, Guy, Hatzidimitriou, Hauser, Haywood, Helmer,
  Helmi, Sarmiento, Hidalgo, Hilger, H{\textbackslash}ladczuk, Hobbs, Holland,
  Huckle, Jardine, Jasniewicz, {Jean-Antoine Piccolo}, {Jim{\'e}nez-Arranz},
  Jorissen, Juaristi~Campillo, Julbe, Karbevska, Kervella, Khanna, Kontizas,
  Kordopatis, Korn, K{\'o}sp{\'a}l, {Kostrzewa-Rutkowska}, Kruszy{\'n}ska, Kun,
  Laizeau, Lambert, Lanza, Lasne, Le~Campion, Lebreton, Lebzelter, Leccia,
  Leclerc, {Lecoeur-Taibi}, Liao, Licata, Lindstr{\o}m, Lister, Livanou, Lobel,
  Lorca, Loup, Madrero~Pardo, Magdaleno~Romeo, Managau, Mann, Manteiga,
  Marchant, Marconi, Marcos, Marcos~Santos, Mar{\'i}n~Pina, Marinoni, Marocco,
  Marshall, Martin~Polo, {Mart{\'i}n-Fleitas}, Marton, Mary, Masip, Massari,
  {Mastrobuono-Battisti}, Mazeh, McMillan, Messina, Michalik, Millar, Mints,
  Molina, Molinaro, Moln{\'a}r, Monari, Mongui{\'o}, Montegriffo, Montero, Mor,
  Mora, Morbidelli, Morel, Morris, Muraveva, Murphy, Musella, Nagy, Noval,
  Oca{\~n}a, Ogden, Ordenovic, Osinde, Pagani, Pagano, Palaversa, Palicio,
  {Pallas-Quintela}, Panahi, {Payne-Wardenaar}, Pe{\~n}alosa~Esteller,
  Penttil{\"a}, Pichon, Piersimoni, Pineau, Plachy, Plum, Poggio, Pr{\v s}a,
  Pulone, Racero, Ragaini, Rainer, Raiteri, Rambaux, Ramos, {Ramos-Lerate},
  Re~Fiorentin, Regibo, Richards, Rios~Diaz, Ripepi, Riva, Rix, Rixon,
  Robichon, Robin, Robin, Roelens, Rogues, Rohrbasser, {Romero-G{\'o}mez},
  Rowell, Royer, Ruz~Mieres, Rybicki, Sadowski, S{\'a}ez~N{\'u}{\~n}ez,
  Sagrist{\`a}~Sell{\'e}s, Sahlmann, Salguero, Samaras, Sanchez~Gimenez, Sanna,
  Santove{\~n}a, Sarasso, Schultheis, Sciacca, Segol, Segovia, S{\'e}gransan,
  Semeux, Shahaf, Siddiqui, Siebert, Siltala, Silvelo, Slezak, Slezak, Smart,
  Snaith, Solano, Solitro, Souami, Souchay, Spagna, Spina, Spoto, Steele,
  Steidelm{\"u}ller, Stephenson, S{\"u}veges, Surdej, Szabados, {Szegedi-Elek},
  Taris, Taylor, Teixeira, Tolomei, Tonello, Torra, Torra, Torralba~Elipe,
  Trabucchi, Tsounis, Turon, Ulla, Unger, Vaillant, {van Dillen}, {van Reeven},
  Vanel, Vecchiato, Viala, Vicente, Voutsinas, Weiler, Wevers, Wyrzykowski,
  Yoldas, Yvard, Zhao, Zorec, Zucker, \& Zwitter}]{GaiaCollaboration2023}
{Gaia Collaboration}, Vallenari, A., Brown, A. G.~A., {et~al.} 2023, A\&A, 674,
  A1, \dodoi{10.1051/0004-6361/202243940}

\bibitem[{Gao {et~al.}(2023)Gao, Gu, Liu, Zhang, Shi, Dou, Li, \&
  Kong}]{Gao2023}
Gao, Y., Gu, Q., Liu, G., {et~al.} 2023, \aap, 677, A179,
  \dodoi{10.1051/0004-6361/202346753}

\bibitem[{{Garay-Solis} {et~al.}(2023){Garay-Solis}, {Barrera-Ballesteros},
  Colombo, S{\'a}nchez, {Lugo-Aranda}, Villanueva, Wong, \&
  Bolatto}]{Garay-Solis2023}
{Garay-Solis}, Y., {Barrera-Ballesteros}, J.~K., Colombo, D., {et~al.} 2023,
  \apj, 952, 122, \dodoi{10.3847/1538-4357/acd781}

\bibitem[{Gardu{\~n}o {et~al.}(2021)Gardu{\~n}o, {Lara-L{\'o}pez},
  {L{\'o}pez-Cruz}, Hopkins, Owers, Pimbblet, \& Holwerda}]{Garduno2021}
Gardu{\~n}o, L.~E., {Lara-L{\'o}pez}, M.~A., {L{\'o}pez-Cruz}, O., {et~al.}
  2021, \mnras, 501, 2969, \dodoi{10.1093/mnras/staa3799}

\bibitem[{Giovanelli {et~al.}(2005)Giovanelli, Haynes, Kent, Perillat,
  Saintonge, Brosch, Catinella, Hoffman, Stierwalt, Spekkens, Lerner, Masters,
  Momjian, Rosenberg, Springob, Boselli, Charmandaris, Darling, Davies, Lambas,
  Gavazzi, Giovanardi, Hardy, Hunt, Iovino, Karachentsev, Karachentseva,
  Koopmann, Marinoni, Minchin, Muller, Putman, Pantoja, Salzer, Scodeggio,
  Skillman, Solanes, Valotto, Van~Driel, \& Van~Zee}]{Giovanelli2005}
Giovanelli, R., Haynes, M.~P., Kent, B.~R., {et~al.} 2005, AJ, 130, 2598,
  \dodoi{10.1086/497431}

\bibitem[{Graziani {et~al.}(2019)Graziani, Courtois, Lavaux, Hoffman, Tully,
  Copin, \& Pomar{\`e}de}]{Graziani2019}
Graziani, R., Courtois, H.~M., Lavaux, G., {et~al.} 2019, \mnras, 488, 5438,
  \dodoi{10.1093/mnras/stz078}

\bibitem[{Guo {et~al.}(2021)Guo, Jones, Wang, \& Lin}]{Guo2021}
Guo, H., Jones, M.~G., Wang, J., \& Lin, L. 2021, ApJ, 918, 53,
  \dodoi{10.3847/1538-4357/ac062e}

\bibitem[{Hani {et~al.}(2020)Hani, Gosain, Ellison, Patton, \&
  Torrey}]{Hani2020}
Hani, M.~H., Gosain, H., Ellison, S.~L., Patton, D.~R., \& Torrey, P. 2020,
  \mnras, 493, 3716, \dodoi{10.1093/mnras/staa459}

\bibitem[{Hani {et~al.}(2018)Hani, Sparre, Ellison, Torrey, \&
  Vogelsberger}]{Hani2018}
Hani, M.~H., Sparre, M., Ellison, S.~L., Torrey, P., \& Vogelsberger, M. 2018,
  \mnras, 475, 1160, \dodoi{10.1093/mnras/stx3252}

\bibitem[{Haynes {et~al.}(2018)Haynes, Giovanelli, Kent, Adams, Balonek, Craig,
  Fertig, Finn, Giovanardi, Hallenbeck, Hess, Hoffman, Huang, Jones, Koopmann,
  Kornreich, Leisman, Miller, Moorman, O'Connor, O'Donoghue, Papastergis,
  Troischt, Stark, \& Xiao}]{Haynes2018}
Haynes, M.~P., Giovanelli, R., Kent, B.~R., {et~al.} 2018, \apj, 861, 49,
  \dodoi{10.3847/1538-4357/aac956}

\bibitem[{He {et~al.}(2024)He, Xu, Lisenfeld, Dai, Fang, Huang, Wang, \&
  Yu}]{He2024}
He, C., Xu, C., Lisenfeld, U., {et~al.} 2024, Res. Astron. Astrophys., 24,
  055005, \dodoi{10.1088/1674-4527/ad392c}

\bibitem[{Huang \& Fan(2022)}]{Huang2022}
Huang, Q., \& Fan, L. 2022, \apjs, 262, 39, \dodoi{10.3847/1538-4365/ac85b1}

\bibitem[{Hunter {et~al.}(2024)Hunter, Elmegreen, \& Madden}]{Hunter2024}
Hunter, D.~A., Elmegreen, B.~G., \& Madden, S.~C. 2024, \araa, 62, 113,
  \dodoi{10.1146/annurev-astro-052722-104109}

\bibitem[{Inoue {et~al.}(2018)Inoue, Hennebelle, Fukui, Matsumoto, Iwasaki, \&
  Inutsuka}]{Inoue2018}
Inoue, T., Hennebelle, P., Fukui, Y., {et~al.} 2018, \pasj, 70, S53,
  \dodoi{10.1093/pasj/psx089}

\bibitem[{Janowiecki {et~al.}(2020)Janowiecki, Catinella, Cortese, Saintonge,
  \& Wang}]{Janowiecki2020}
Janowiecki, S., Catinella, B., Cortese, L., Saintonge, A., \& Wang, J. 2020,
  \mnras, 493, 1982, \dodoi{10.1093/mnras/staa178}

\bibitem[{Jarrett {et~al.}(2011)Jarrett, Cohen, Masci, Wright, Stern, Benford,
  Blain, Carey, Cutri, Eisenhardt, Lonsdale, Mainzer, Marsh, Padgett, Petty,
  Ressler, Skrutskie, Stanford, Surace, Tsai, Wheelock, \& Yan}]{Jarrett2011}
Jarrett, T.~H., Cohen, M., Masci, F., {et~al.} 2011, \apj, 735, 112,
  \dodoi{10.1088/0004-637X/735/2/112}

\bibitem[{Jarrett {et~al.}(2013)Jarrett, Masci, Tsai, Petty, Cluver, Assef,
  Benford, Blain, Bridge, Donoso, Eisenhardt, Koribalski, Lake, Neill, Seibert,
  Sheth, Stanford, \& Wright}]{Jarrett2013}
Jarrett, T.~H., Masci, F., Tsai, C.~W., {et~al.} 2013, \aj, 145, 6,
  \dodoi{10.1088/0004-6256/145/1/6}

\bibitem[{Jog \& Solomon(1992)}]{Jog1992}
Jog, C.~J., \& Solomon, P.~M. 1992, ApJ, 387, 152, \dodoi{10.1086/171067}

\bibitem[{{Kado-Fong} {et~al.}(2020){Kado-Fong}, Greene, Greco, Beaton,
  Goulding, Johnson, \& Komiyama}]{Kado-Fong2020}
{Kado-Fong}, E., Greene, J.~E., Greco, J.~P., {et~al.} 2020, \aj, 159, 103,
  \dodoi{10.3847/1538-3881/ab6ef3}

\bibitem[{{Kado-Fong} {et~al.}(2024){Kado-Fong}, Robinson, Nyland, Greene,
  Suess, Stierwalt, \& Beaton}]{Kado-Fong2024}
{Kado-Fong}, E., Robinson, A., Nyland, K., {et~al.} 2024, \apj, 963, 37,
  \dodoi{10.3847/1538-4357/ad18cb}

\bibitem[{Keel {et~al.}(1985)Keel, Kennicutt, Hummel, \& {van der
  Hulst}}]{Keel1985}
Keel, W.~C., Kennicutt, Jr., R.~C., Hummel, E., \& {van der Hulst}, J.~M. 1985,
  Astron. J., 90, 708, \dodoi{10.1086/113779}

\bibitem[{Kennicutt \& Evans(2012)}]{Kennicutt2012}
Kennicutt, R.~C., \& Evans, N.~J. 2012, \araa, 50, 531,
  \dodoi{10.1146/annurev-astro-081811-125610}

\bibitem[{Kewley {et~al.}(2006)Kewley, Geller, \& Barton}]{Kewley2006}
Kewley, L.~J., Geller, M.~J., \& Barton, E.~J. 2006, AJ, 131, 2004,
  \dodoi{10.1086/500295}

\bibitem[{Kewley {et~al.}(2010)Kewley, Rupke, Zahid, Geller, \&
  Barton}]{Kewley2010}
Kewley, L.~J., Rupke, D., Zahid, H.~J., Geller, M.~J., \& Barton, E.~J. 2010,
  \apjl, 721, L48, \dodoi{10.1088/2041-8205/721/1/L48}

\bibitem[{Kim {et~al.}(2023)Kim, Oh, Wang, {Staveley-Smith}, Koribalski, Kim,
  Park, Kim, Spekkens, Westmeier, Wong, Meurer, Kamphuis, Catinella, McQuinn,
  Bigiel, Holwerda, Rhee, {Lee-Waddell}, Deg, {Verdes-Montenegro}, For, Madrid,
  D{\'e}nes, \& Elagali}]{Kim2023}
Kim, S.-J., Oh, S.-H., Wang, J., {et~al.} 2023, \mnras, 519, 318,
  \dodoi{10.1093/mnras/stac3480}

\bibitem[{Koribalski {et~al.}(2020)Koribalski, {Staveley-Smith}, Westmeier,
  Serra, Spekkens, Wong, {Lee-Waddell}, Lagos, Obreschkow, {Ryan-Weber}, Zwaan,
  Kilborn, Bekiaris, Bekki, Bigiel, Boselli, Bosma, Catinella, Chauhan, Cluver,
  Colless, Courtois, Crain, {de Blok}, D{\'e}nes, Duffy, Elagali, Fluke, For,
  Heald, Henning, Hess, Holwerda, Howlett, Jarrett, Jones, Jones, J{\'o}zsa,
  Jurek, J{\"u}tte, Kamphuis, Karachentsev, Kerp, Kleiner, {Kraan-Korteweg},
  {L{\'o}pez-S{\'a}nchez}, Madrid, Meyer, Mould, Murugeshan, Norris, Oh,
  Oosterloo, Popping, Putman, Reynolds, Rhee, Robotham, Ryder, Schr{\"o}der,
  Shao, Stevens, Taylor, {van{\^A} der Hulst}, {Verdes-Montenegro}, Wakker,
  Wang, Whiting, Winkel, \& Wolf}]{Koribalski2020}
Koribalski, B.~S., {Staveley-Smith}, L., Westmeier, T., {et~al.} 2020, \apss,
  365, 118, \dodoi{10.1007/s10509-020-03831-4}

\bibitem[{Kourkchi {et~al.}(2020)Kourkchi, Courtois, Graziani, Hoffman,
  Pomar{\`e}de, Shaya, \& Tully}]{Kourkchi2020}
Kourkchi, E., Courtois, H.~M., Graziani, R., {et~al.} 2020, AJ, 159, 67,
  \dodoi{10.3847/1538-3881/ab620e}

\bibitem[{Kourkchi \& Tully(2017)}]{Kourkchi2017}
Kourkchi, E., \& Tully, R.~B. 2017, \apj, 843, 16,
  \dodoi{10.3847/1538-4357/aa76db}

\bibitem[{Lang(2014)}]{Lang2014}
Lang, D. 2014, \aj, 147, 108, \dodoi{10.1088/0004-6256/147/5/108}

\bibitem[{Lang {et~al.}(2016)Lang, Hogg, \& Mykytyn}]{Lang2016}
Lang, D., Hogg, D.~W., \& Mykytyn, D. 2016, The {{Tractor}}: {{Probabilistic
  Astronomical Source Detection}} and {{Measurement}}

\bibitem[{Larson {et~al.}(2016)Larson, Sanders, Barnes, Ishida, Evans, U,
  Mazzarella, Kim, Privon, Mirabel, \& Flewelling}]{Larson2016}
Larson, K.~L., Sanders, D.~B., Barnes, J.~E., {et~al.} 2016, \apj, 825, 128,
  \dodoi{10.3847/0004-637X/825/2/128}

\bibitem[{Lee {et~al.}(2022)Lee, Wang, Chung, Ho, Wang, Michiyama, Molina, Kim,
  Shao, Kilborn, Wang, Lin, Kim, Catinella, Cortese, Deg, Denes, Elagali, For,
  Kleiner, Koribalski, {Lee-Waddell}, Rhee, Spekkens, Westmeier, Wong, Bigiel,
  Bosma, Holwerda, {van der Hulst}, Roychowdhury, {Verdes-Montenegro}, \&
  Zwaan}]{Lee2022}
Lee, B., Wang, J., Chung, A., {et~al.} 2022, \apjs, 262, 31,
  \dodoi{10.3847/1538-4365/ac7eba}

\bibitem[{Li {et~al.}(2023{\natexlab{a}})Li, Ho, \& Shangguan}]{Li2023a}
Li, Y.~A., Ho, L.~C., \& Shangguan, J. 2023{\natexlab{a}}, \apj, 953, 91,
  \dodoi{10.3847/1538-4357/acdddb}

\bibitem[{Li {et~al.}(2023{\natexlab{b}})Li, Ho, Shangguan, Zhuang, \&
  Li}]{Li2023b}
Li, Y.~A., Ho, L.~C., Shangguan, J., Zhuang, M.-Y., \& Li, R.
  2023{\natexlab{b}}, ApJS, 267, 17, \dodoi{10.3847/1538-4365/acd4b5}

\bibitem[{Lin {et~al.}(2023)Lin, Wang, Kilborn, Peng, Cortese, Boselli, Liang,
  Lee, Yang, Catinella, Deg, D{\'e}nes, Elagali, Kamphuis, Koribalski,
  {Lee-Waddell}, Rhee, Shao, Spekkens, {Staveley-Smith}, Westmeier, Wong,
  Bekki, Bosma, Du, Ho, Madrid, {Verdes-Montenegro}, Wang, \& Wang}]{Lin2023}
Lin, X., Wang, J., Kilborn, V., {et~al.} 2023, ApJ, 956, 148,
  \dodoi{10.3847/1538-4357/accea2}

\bibitem[{Lisenfeld {et~al.}(2019)Lisenfeld, Xu, Gao, Domingue, Cao, Yun, \&
  Zuo}]{Lisenfeld2019}
Lisenfeld, U., Xu, C.~K., Gao, Y., {et~al.} 2019, A\&A, 627, A107,
  \dodoi{10.1051/0004-6361/201935536}

\bibitem[{Martig {et~al.}(2009)Martig, Bournaud, Teyssier, \&
  Dekel}]{Martig2009}
Martig, M., Bournaud, F., Teyssier, R., \& Dekel, A. 2009, \apj, 707, 250,
  \dodoi{10.1088/0004-637X/707/1/250}

\bibitem[{Martin {et~al.}(2005)Martin, Fanson, Schiminovich, Morrissey,
  Friedman, Barlow, Conrow, Grange, Jelinsky, Milliard, Siegmund, Bianchi,
  Byun, Donas, Forster, Heckman, Lee, Madore, Malina, Neff, Rich, Small,
  Surber, Szalay, Welsh, \& Wyder}]{Martin2005}
Martin, D.~C., Fanson, J., Schiminovich, D., {et~al.} 2005, \apjl, 619, L1,
  \dodoi{10.1086/426387}

\bibitem[{Masters {et~al.}(2012)Masters, Nichol, Haynes, Keel, Lintott,
  Simmons, Skibba, Bamford, Giovanelli, \& Schawinski}]{Masters2012}
Masters, K.~L., Nichol, R.~C., Haynes, M.~P., {et~al.} 2012, \mnras, 424, 2180,
  \dodoi{10.1111/j.1365-2966.2012.21377.x}

\bibitem[{McConnachie(2012)}]{McConnachie2012}
McConnachie, A.~W. 2012, \aj, 144, 4, \dodoi{10.1088/0004-6256/144/1/4}

\bibitem[{McGaugh {et~al.}(2000)McGaugh, Schombert, Bothun, \& {de
  Blok}}]{McGaugh2000}
McGaugh, S.~S., Schombert, J.~M., Bothun, G.~D., \& {de Blok}, W. J.~G. 2000,
  \apjl, 533, L99, \dodoi{10.1086/312628}

\bibitem[{Mirabel \& Sanders(1989)}]{Mirabel1989}
Mirabel, I.~F., \& Sanders, D.~B. 1989, \apj, 340, L53, \dodoi{10.1086/185437}

\bibitem[{Moon {et~al.}(2019)Moon, An, \& Yoon}]{Moon2019}
Moon, J.-S., An, S.-H., \& Yoon, S.-J. 2019, \apj, 882, 14,
  \dodoi{10.3847/1538-4357/ab3401}

\bibitem[{Moreno {et~al.}(2013)Moreno, Bluck, Ellison, Patton, Torrey, \&
  Moster}]{Moreno2013}
Moreno, J., Bluck, A. F.~L., Ellison, S.~L., {et~al.} 2013, MNRAS, 436, 1765,
  \dodoi{10.1093/mnras/stt1694}

\bibitem[{Moreno {et~al.}(2019)Moreno, Torrey, Ellison, Patton, Hopkins, Bueno,
  Hayward, Narayanan, Kere{\v s}, Bluck, \& Hernquist}]{Moreno2019}
Moreno, J., Torrey, P., Ellison, S.~L., {et~al.} 2019, \mnras, 485, 1320,
  \dodoi{10.1093/mnras/stz417}

\bibitem[{Murugeshan {et~al.}(2024)Murugeshan, Deg, Westmeier, Shen, For,
  Spekkens, Wong, {Staveley-Smith}, Catinella, {Lee-Waddell}, D{\'e}nes, Rhee,
  Cortese, Goliath, Halloran, {van der Hulst}, Kamphuis, Koribalski,
  {Kraan-Korteweg}, Lelli, Venkataraman, {Verdes-Montenegro}, \&
  Yu}]{Murugeshan2024}
Murugeshan, C., Deg, N., Westmeier, T., {et~al.} 2024, {{WALLABY Pilot
  Survey}}: {{Public}} Data Release of {\textasciitilde}1800 {{HI}} Sources and
  High-Resolution Cut-Outs from {{Pilot Survey Phase}} 2,  arXiv.
\newblock \doarXiv{2409.13130}

\bibitem[{O'Beirne {et~al.}(2024)O'Beirne, {Staveley-Smith}, Wong, Westmeier,
  Batten, Kilborn, {Lee-Waddell}, Mancera~Pi{\~n}a, Rom{\'a}n,
  {Verdes-Montenegro}, Catinella, Cortese, Deg, D{\'e}nes, For, Kamphuis,
  Koribalski, Murugeshan, Rhee, Spekkens, Wang, Bekki, \&
  {L{\'p}pez-S{\'a}nchez}}]{OBeirne2024}
O'Beirne, T., {Staveley-Smith}, L., Wong, O.~I., {et~al.} 2024, \mnras, 528,
  4010, \dodoi{10.1093/mnras/stae215}

\bibitem[{Pan {et~al.}(2018)Pan, Lin, Hsieh, Xiao, Gao, Ellison, Scudder,
  {Barrera-Ballesteros}, Yuan, Saintonge, Wilson, Hwang, De~Looze, Gao, Ho,
  Brinks, Mok, Brown, Davis, Williams, Chung, Parsons, Bureau, Sargent, Chung,
  Kim, Liu, Micha{\textbackslash}lowski, \& Tosaki}]{Pan2018}
Pan, H.-A., Lin, L., Hsieh, B.-C., {et~al.} 2018, ApJ, 868, 132,
  \dodoi{10.3847/1538-4357/aaeb92}

\bibitem[{Pan {et~al.}(2019)Pan, Lin, Hsieh, {Barrera-Ballesteros},
  S{\'a}nchez, Hsu, Keenan, Tissera, Boquien, Dai, Knapen, Riffel,
  {Argudo-Fern{\'a}ndez}, Xiao, \& Yuan}]{Pan2019}
---. 2019, \apj, 881, 119, \dodoi{10.3847/1538-4357/ab311c}

\bibitem[{Park \& Choi(2009)}]{Park2009}
Park, C., \& Choi, Y.-Y. 2009, \apj, 691, 1828,
  \dodoi{10.1088/0004-637X/691/2/1828}

\bibitem[{Patton {et~al.}(2013)Patton, Torrey, Ellison, Mendel, \&
  Scudder}]{Patton2013}
Patton, D.~R., Torrey, P., Ellison, S.~L., Mendel, J.~T., \& Scudder, J.~M.
  2013, \mnras, 433, L59, \dodoi{10.1093/mnrasl/slt058}

\bibitem[{Patton {et~al.}(2020)Patton, Wilson, Metrow, Ellison, Torrey, Brown,
  Hani, McAlpine, Moreno, \& Woo}]{Patton2020}
Patton, D.~R., Wilson, K.~D., Metrow, C.~J., {et~al.} 2020, \mnras, 494, 4969,
  \dodoi{10.1093/mnras/staa913}

\bibitem[{Paudel {et~al.}(2018)Paudel, Smith, Yoon, {Calder{\'o}n-Castillo}, \&
  Duc}]{Paudel2018}
Paudel, S., Smith, R., Yoon, S.~J., {Calder{\'o}n-Castillo}, P., \& Duc, P.-A.
  2018, \apjs, 237, 36, \dodoi{10.3847/1538-4365/aad555}

\bibitem[{Pawlik {et~al.}(2016)Pawlik, Wild, Walcher, Johansson, Villforth,
  Rowlands, {Mendez-Abreu}, \& Hewlett}]{Pawlik2016}
Pawlik, M.~M., Wild, V., Walcher, C.~J., {et~al.} 2016, \mnras, 456, 3032,
  \dodoi{10.1093/mnras/stv2878}

\bibitem[{Pearson {et~al.}(2016)Pearson, Besla, Putman, Lutz, Fern{\'a}ndez,
  Stierwalt, Patton, Kim, Kallivayalil, Johnson, \& Sung}]{Pearson2016}
Pearson, S., Besla, G., Putman, M.~E., {et~al.} 2016, \mnras, 459, 1827,
  \dodoi{10.1093/mnras/stw757}

\bibitem[{Pearson {et~al.}(2018)Pearson, Privon, Besla, Putman,
  {Mart{\'i}nez-Delgado}, Johnston, Gabany, Patton, \&
  Kallivayalil}]{Pearson2018}
Pearson, S., Privon, G.~C., Besla, G., {et~al.} 2018, \mnras, 480, 3069,
  \dodoi{10.1093/mnras/sty2052}

\bibitem[{{P{\'e}rez-D{\'i}az} {et~al.}(2024){P{\'e}rez-D{\'i}az},
  {P{\'e}rez-Montero}, {Fern{\'a}ndez-Ontiveros}, V{\'i}lchez, \&
  Amor{\'i}n}]{Perez-Diaz2024}
{P{\'e}rez-D{\'i}az}, B., {P{\'e}rez-Montero}, E., {Fern{\'a}ndez-Ontiveros},
  J.~A., V{\'i}lchez, J.~M., \& Amor{\'i}n, R. 2024, \nat, 8, 368,
  \dodoi{10.1038/s41550-023-02171-x}

\bibitem[{Petersson {et~al.}(2022)Petersson, Renaud, Agertz, Dekel, \&
  Duc}]{Petersson2022}
Petersson, J., Renaud, F., Agertz, O., Dekel, A., \& Duc, P.-A. 2022, MNRAS,
  518, 3261, \dodoi{10.1093/mnras/stac3136}

\bibitem[{Piffaretti {et~al.}(2011)Piffaretti, Arnaud, Pratt, Pointecouteau, \&
  Melin}]{Piffaretti2011}
Piffaretti, R., Arnaud, M., Pratt, G.~W., Pointecouteau, E., \& Melin, J.~B.
  2011, \aap, 534, A109, \dodoi{10.1051/0004-6361/201015377}

\bibitem[{Pillepich {et~al.}(2019)Pillepich, Nelson, Springel, Pakmor, Torrey,
  Weinberger, Vogelsberger, Marinacci, Genel, {van~der~Wel}, \&
  Hernquist}]{Pillepich2019}
Pillepich, A., Nelson, D., Springel, V., {et~al.} 2019, MNRAS, 490, 3196,
  \dodoi{10.1093/mnras/stz2338}

\bibitem[{Punzo {et~al.}(2017)Punzo, {van der Hulst}, Roerdink,
  {Fillion-Robin}, \& Yu}]{Punzo2017}
Punzo, D., {van der Hulst}, J.~M., Roerdink, J. B. T.~M., {Fillion-Robin},
  J.~C., \& Yu, L. 2017, Astron. Comput., 19, 45,
  \dodoi{10.1016/j.ascom.2017.03.004}

\bibitem[{Quai {et~al.}(2023)Quai, {Byrne-Mamahit}, Ellison, Patton, \&
  Hani}]{Quai2023}
Quai, S., {Byrne-Mamahit}, S., Ellison, S.~L., Patton, D.~R., \& Hani, M.~H.
  2023, \mnras, 519, 2119, \dodoi{10.1093/mnras/stac3713}

\bibitem[{Reiprich \& B{\"o}hringer(2002)}]{Reiprich2002}
Reiprich, T.~H., \& B{\"o}hringer, H. 2002, \apj, 567, 716,
  \dodoi{10.1086/338753}

\bibitem[{Renaud {et~al.}(2022)Renaud, Segovia~Otero, \& Agertz}]{Renaud2022}
Renaud, F., Segovia~Otero, {\'A}., \& Agertz, O. 2022, \mnras, 516, 4922,
  \dodoi{10.1093/mnras/stac2557}

\bibitem[{Reynolds {et~al.}(2022)Reynolds, Catinella, Cortese, Westmeier,
  Meurer, Shao, Obreschkow, Rom{\'a}n, {Verdes-Montenegro}, Deg, D{\'e}nes,
  For, Kleiner, Koribalski, {Lee-Waddell}, Murugeshan, Oh, Rhee, Spekkens,
  {Staveley-Smith}, Stevens, {van der Hulst}, Wang, Wong, Holwerda, Bosma,
  Madrid, \& Bekki}]{Reynolds2022}
Reynolds, T.~N., Catinella, B., Cortese, L., {et~al.} 2022, \mnras, 510, 1716,
  \dodoi{10.1093/mnras/stab3522}

\bibitem[{Saintonge \& Catinella(2022)}]{Saintonge2022}
Saintonge, A., \& Catinella, B. 2022, \araa, 60, 319,
  \dodoi{10.1146/annurev-astro-021022-043545}

\bibitem[{Saintonge {et~al.}(2017)Saintonge, Catinella, Tacconi, Kauffmann,
  Genzel, Cortese, Dav{\'e}, Fletcher, {Graci{\'a}-Carpio}, Kramer, Heckman,
  Janowiecki, Lutz, Rosario, Schiminovich, Schuster, Wang, Wuyts, Borthakur,
  Lamperti, \& {Roberts-Borsani}}]{Saintonge2017}
Saintonge, A., Catinella, B., Tacconi, L.~J., {et~al.} 2017, ApJS, 233, 22,
  \dodoi{10.3847/1538-4365/aa97e0}

\bibitem[{Salim {et~al.}(2018)Salim, Boquien, \& Lee}]{Salim2018}
Salim, S., Boquien, M., \& Lee, J.~C. 2018, \apj, 859, 11,
  \dodoi{10.3847/1538-4357/aabf3c}

\bibitem[{Salim {et~al.}(2016)Salim, Lee, Janowiecki, {da Cunha}, Dickinson,
  Boquien, Burgarella, Salzer, \& Charlot}]{Salim2016}
Salim, S., Lee, J.~C., Janowiecki, S., {et~al.} 2016, \apjs, 227, 2,
  \dodoi{10.3847/0067-0049/227/1/2}

\bibitem[{Satyapal {et~al.}(2014)Satyapal, Ellison, McAlpine, Hickox, Patton,
  \& Mendel}]{Satyapal2014}
Satyapal, S., Ellison, S.~L., McAlpine, W., {et~al.} 2014, \mnras, 441, 1297,
  \dodoi{10.1093/mnras/stu650}

\bibitem[{Scudder {et~al.}(2015)Scudder, Ellison, Momjian, Rosenberg, Torrey,
  Patton, Fertig, \& Mendel}]{Scudder2015}
Scudder, J.~M., Ellison, S.~L., Momjian, E., {et~al.} 2015, \mnras, 449, 3719,
  \dodoi{10.1093/mnras/stv588}

\bibitem[{Serra {et~al.}(2012)Serra, Oosterloo, Morganti, Alatalo, Blitz, Bois,
  Bournaud, Bureau, Cappellari, Crocker, Davies, Davis, {de Zeeuw}, Duc,
  Emsellem, Khochfar, Krajnovi{\'c}, Kuntschner, Lablanche, McDermid, Naab,
  Sarzi, Scott, Trager, Weijmans, \& Young}]{Serra2012}
Serra, P., Oosterloo, T., Morganti, R., {et~al.} 2012, \mnras, 422, 1835,
  \dodoi{10.1111/j.1365-2966.2012.20219.x}

\bibitem[{Serra {et~al.}(2015)Serra, Westmeier, Giese, Jurek, Fl{\"o}er,
  Popping, Winkel, {van der Hulst}, Meyer, Koribalski, {Staveley-Smith}, \&
  Courtois}]{Serra2015}
Serra, P., Westmeier, T., Giese, N., {et~al.} 2015, \mnras, 448, 1922,
  \dodoi{10.1093/mnras/stv079}

\bibitem[{Shah {et~al.}(2022)Shah, Kartaltepe, Magagnoli, Cox, Wetherell,
  Vanderhoof, Cooke, Calabro, Chartab, Conselice, Croton, De~La~Vega, Hathi,
  Ilbert, Inami, Kocevski, Koekemoer, Lemaux, Lubin, Mantha, Marchesi, Martig,
  Moreno, Pampliega, Patton, Salvato, \& Treister}]{Shah2022}
Shah, E.~A., Kartaltepe, J.~S., Magagnoli, C.~T., {et~al.} 2022, ApJ, 940, 4,
  \dodoi{10.3847/1538-4357/ac96eb}

\bibitem[{Silverman {et~al.}(2011)Silverman, Kampczyk, Jahnke, Andrae, Lilly,
  Elvis, Civano, Mainieri, Vignali, Zamorani, Nair, Le~F{\`e}vre, De~Ravel,
  Bardelli, Bongiorno, Bolzonella, Cappi, Caputi, Carollo, Contini, Coppa,
  Cucciati, De~La~Torre, Franzetti, Garilli, Halliday, Hasinger, Iovino,
  Knobel, Koekemoer, Kova{\v c}, Lamareille, Le~Borgne, Le~Brun, Maier,
  Mignoli, Pello, {P{\'e}rez-Montero}, Ricciardelli, Peng, Scodeggio, Tanaka,
  Tasca, Tresse, Vergani, Zucca, Brusa, Cappelluti, Comastri, Finoguenov, Fu,
  Gilli, Hao, Ho, \& Salvato}]{Silverman2011}
Silverman, J.~D., Kampczyk, P., Jahnke, K., {et~al.} 2011, ApJ, 743, 2,
  \dodoi{10.1088/0004-637X/743/1/2}

\bibitem[{Sparre {et~al.}(2022)Sparre, Whittingham, Damle, Hani, Richter,
  Ellison, Pfrommer, \& Vogelsberger}]{Sparre2022}
Sparre, M., Whittingham, J., Damle, M., {et~al.} 2022, \mnras, 509, 2720,
  \dodoi{10.1093/mnras/stab3171}

\bibitem[{Spilker {et~al.}(2022)Spilker, Suess, Setton, Bezanson, Feldmann,
  Greene, Kriek, Lower, Narayanan, \& Verrico}]{Spilker2022}
Spilker, J.~S., Suess, K.~A., Setton, D.~J., {et~al.} 2022, \apjl, 936, L11,
  \dodoi{10.3847/2041-8213/ac75ea}

\bibitem[{Stierwalt {et~al.}(2015)Stierwalt, Besla, Patton, Johnson,
  Kallivayalil, Putman, Privon, \& Ross}]{Stierwalt2015}
Stierwalt, S., Besla, G., Patton, D., {et~al.} 2015, ApJ, 805, 2,
  \dodoi{10.1088/0004-637X/805/1/2}

\bibitem[{Subramanian {et~al.}(2024)Subramanian, Mondal, \&
  Kalari}]{Subramanian2024}
Subramanian, S., Mondal, C., \& Kalari, V. 2024, A\&A, 681, A8,
  \dodoi{10.1051/0004-6361/202346536}

\bibitem[{{The Astropy Collaboration} {et~al.}(2013){The Astropy
  Collaboration}, Robitaille, Tollerud, Greenfield, Droettboom, Bray, Aldcroft,
  Davis, Ginsburg, {Price-Whelan}, Kerzendorf, Conley, Crighton, Barbary, Muna,
  Ferguson, Grollier, Parikh, Nair, G{\"u}nther, Deil, Woillez, Conseil,
  Kramer, Turner, Singer, Fox, Weaver, Zabalza, Edwards, Azalee~Bostroem,
  Burke, Casey, Crawford, Dencheva, Ely, Jenness, Labrie, Lim, Pierfederici,
  Pontzen, Ptak, Refsdal, Servillat, \&
  Streicher}]{TheAstropyCollaboration2013}
{The Astropy Collaboration}, Robitaille, T.~P., Tollerud, E.~J., {et~al.} 2013,
  A\&A, 558, A33, \dodoi{10.1051/0004-6361/201322068}

\bibitem[{{The Astropy Collaboration} {et~al.}(2022){The Astropy
  Collaboration}, {Price-Whelan}, Lim, Earl, Starkman, Bradley, Shupe, Patil,
  Corrales, Brasseur, N{\"o}the, Donath, Tollerud, Morris, Ginsburg, Vaher,
  Weaver, Tocknell, Jamieson, Van~Kerkwijk, Robitaille, Merry, Bachetti,
  G{\"u}nther, {Paper Authors}, Aldcroft, {Alvarado-Montes}, Archibald,
  B{\'o}di, Bapat, Barentsen, Baz{\'a}n, Biswas, Boquien, Burke, Cara, Cara,
  Conroy, Conseil, Craig, Cross, Cruz, D'Eugenio, Dencheva, Devillepoix,
  Dietrich, Eigenbrot, Erben, Ferreira, {Foreman-Mackey}, Fox, Freij, Garg,
  Geda, Glattly, Gondhalekar, Gordon, Grant, Greenfield, Groener, Guest,
  Gurovich, Handberg, Hart, {Hatfield-Dodds}, Homeier, Hosseinzadeh, Jenness,
  Jones, Joseph, Kalmbach, Karamehmetoglu, Ka{\l}uszy{\'n}ski, Kelley, Kern,
  Kerzendorf, Koch, Kulumani, Lee, Ly, Ma, MacBride, Maljaars, Muna, Murphy,
  Norman, O'Steen, Oman, Pacifici, Pascual, {Pascual-Granado}, Patil, Perren,
  Pickering, Rastogi, Roulston, Ryan, Rykoff, Sabater, Sakurikar, Salgado,
  Sanghi, Saunders, Savchenko, Schwardt, {Seifert-Eckert}, Shih, Jain, Shukla,
  Sick, Simpson, Singanamalla, Singer, Singhal, Sinha, Sip{\H o}cz, Spitler,
  Stansby, Streicher, {\v S}umak, Swinbank, Taranu, Tewary, Tremblay,
  {Val-Borro}, Van~Kooten, Vasovi{\'c}, Verma, De~Miranda~Cardoso, Williams,
  Wilson, Winkel, {Wood-Vasey}, Xue, Yoachim, Zhang, Zonca, \& {Astropy Project
  Contributors}}]{TheAstropyCollaboration2022}
{The Astropy Collaboration}, {Price-Whelan}, A.~M., Lim, P.~L., {et~al.} 2022,
  ApJ, 935, 167, \dodoi{10.3847/1538-4357/ac7c74}

\bibitem[{Thorp {et~al.}(2019)Thorp, Ellison, Simard, S{\'a}nchez, \&
  Antonio}]{Thorp2019}
Thorp, M.~D., Ellison, S.~L., Simard, L., S{\'a}nchez, S.~F., \& Antonio, B.
  2019, \mnras, 482, L55, \dodoi{10.1093/mnrasl/sly185}

\bibitem[{Toomre \& Toomre(1972)}]{Toomre1972}
Toomre, A., \& Toomre, J. 1972, \apj, 178, 623, \dodoi{10.1086/151823}

\bibitem[{Tully(2015)}]{Tully2015}
Tully, R.~B. 2015, \aj, 149, 171, \dodoi{10.1088/0004-6256/149/5/171}

\bibitem[{Tully {et~al.}(2009)Tully, Rizzi, Shaya, Courtois, Makarov, \&
  Jacobs}]{Tully2009}
Tully, R.~B., Rizzi, L., Shaya, E.~J., {et~al.} 2009, \aj, 138, 323,
  \dodoi{10.1088/0004-6256/138/2/323}

\bibitem[{Tully {et~al.}(2023)Tully, Kourkchi, Courtois, Anand, Blakeslee,
  Brout, {de Jaeger}, Dupuy, Guinet, Howlett, Jensen, Pomar{\`e}de, Rizzi,
  Rubin, Said, Scolnic, \& Stahl}]{Tully2023}
Tully, R.~B., Kourkchi, E., Courtois, H.~M., {et~al.} 2023, \apj, 944, 94,
  \dodoi{10.3847/1538-4357/ac94d8}

\bibitem[{{van der Walt} {et~al.}(2014){van der Walt}, Sch{\"o}nberger,
  {Nunez-Iglesias}, Boulogne, Warner, Yager, Gouillart, \& Yu}]{vanderWalt2014}
{van der Walt}, S., Sch{\"o}nberger, J.~L., {Nunez-Iglesias}, J., {et~al.}
  2014, PeerJ, 2, e453, \dodoi{10.7717/peerj.453}

\bibitem[{Wang {et~al.}(2020)Wang, Catinella, Saintonge, Pan, Serra, \&
  Shao}]{Wang2020}
Wang, J., Catinella, B., Saintonge, A., {et~al.} 2020, ApJ, 890, 63,
  \dodoi{10.3847/1538-4357/ab68dd}

\bibitem[{Wang {et~al.}(2016)Wang, Koribalski, Serra, {van der Hulst},
  Roychowdhury, Kamphuis, \& Chengalur}]{Wang2016}
Wang, J., Koribalski, B.~S., Serra, P., {et~al.} 2016, \mnras, 460, 2143,
  \dodoi{10.1093/mnras/stw1099}

\bibitem[{Wang {et~al.}(2017)Wang, Koribalski, Jarrett, Kamphuis, Li, Ho,
  Westmeier, Shao, Lagos, Wong, Serra, {Staveley-Smith}, J{\'o}zsa, {van der
  Hulst}, \& {L{\'o}pez-S{\'a}nchez}}]{Wang2017}
Wang, J., Koribalski, B.~S., Jarrett, T.~H., {et~al.} 2017, \mnras, 472, 3029,
  \dodoi{10.1093/mnras/stx2073}

\bibitem[{Wang {et~al.}(2021)Wang, {Staveley-Smith}, Westmeier, Catinella,
  Shao, Reynolds, For, Lee, Liang, Wang, Elagali, D{\'e}nes, Kleiner,
  Koribalski, {Lee-Waddell}, Oh, Rhee, Serra, Spekkens, Wong, Bekki, Bigiel,
  Courtois, Hess, Holwerda, McQuinn, {Pandey-Pommier}, {van der Hulst}, \&
  {Verdes-Montenegro}}]{Wang2021}
Wang, J., {Staveley-Smith}, L., Westmeier, T., {et~al.} 2021, \apj, 915, 70,
  \dodoi{10.3847/1538-4357/abfc52}

\bibitem[{Wang {et~al.}(2023)Wang, Yang, Oh, {Staveley-Smith}, Wang, Wang,
  Hess, Ho, Hou, Jing, Kamphuis, Li, Lin, Liu, Shao, Wang, \& Zhu}]{Wang2023a}
Wang, J., Yang, D., Oh, S.~H., {et~al.} 2023, \apj, 944, 102,
  \dodoi{10.3847/1538-4357/acafe8}

\bibitem[{Wang {et~al.}(2022)Wang, Wang, For, Lee, Reynolds, Lin,
  {Staveley-Smith}, Shao, Wong, Catinella, Serra, {Verdes-Montenegro},
  Westmeier, {Lee-Waddell}, Koribalski, Murugeshan, Elagali, Kleiner, Rhee,
  Bigiel, Bosma, Holwerda, Oh, \& Spekkens}]{Wang2022}
Wang, S., Wang, J., For, B.-Q., {et~al.} 2022, \apj, 927, 66,
  \dodoi{10.3847/1538-4357/ac4270}

\bibitem[{Westmeier {et~al.}(2021)Westmeier, Kitaeff, Pallot, Serra, {van der
  Hulst}, Jurek, Elagali, For, Kleiner, Koribalski, {Lee-Waddell}, Mould,
  Reynolds, Rhee, \& {Staveley-Smith}}]{Westmeier2021}
Westmeier, T., Kitaeff, S., Pallot, D., {et~al.} 2021, \mnras, 506, 3962,
  \dodoi{10.1093/mnras/stab1881}

\bibitem[{Westmeier {et~al.}(2022)Westmeier, Deg, Spekkens, Reynolds, Shen,
  Gaudet, Goliath, Huynh, Venkataraman, Lin, O'Beirne, Catinella, Cortese,
  D{\'e}nes, Elagali, For, J{\'o}zsa, Howlett, {van der Hulst}, Jurek,
  Kamphuis, Kilborn, Kleiner, Koribalski, {Lee-Waddell}, Murugeshan, Rhee,
  Serra, Shao, {Staveley-Smith}, Wang, Wong, Zwaan, Allison, Anderson, Ball,
  Bock, Brodrick, Bunton, Cooray, Gupta, Hayman, Mahony, Moss, Ng, Pearce,
  Raja, Roxby, Voronkov, Warhurst, Courtois, \& Said}]{Westmeier2022}
Westmeier, T., Deg, N., Spekkens, K., {et~al.} 2022, \pasa, 39, e058,
  \dodoi{10.1017/pasa.2022.50}

\bibitem[{Wong {et~al.}(2021)Wong, Stevens, For, Westmeier, Dixon, Oh,
  J{\'o}zsa, Reynolds, {Lee-Waddell}, Rom{\'a}n, {Verdes-Montenegro}, Courtois,
  Pomar{\`e}de, Murugeshan, Whiting, Bekki, Bigiel, Bosma, Catinella,
  D{\'e}nes, Elagali, Holwerda, Kamphuis, Kilborn, Kleiner, Koribalski, Lelli,
  Madrid, McQuinn, Popping, Rhee, Roychowdhury, Scott, Sengupta, Spekkens,
  {Staveley-Smith}, \& Wakker}]{Wong2021}
Wong, O.~I., Stevens, A. R.~H., For, B.~Q., {et~al.} 2021, \mnras, 507, 2905,
  \dodoi{10.1093/mnras/stab2262}

\bibitem[{Woods {et~al.}(2006)Woods, Geller, \& Barton}]{Woods2006}
Woods, D.~F., Geller, M.~J., \& Barton, E.~J. 2006, AJ, 132, 197,
  \dodoi{10.1086/504834}

\bibitem[{Wright {et~al.}(2010)Wright, Eisenhardt, Mainzer, Ressler, Cutri,
  Jarrett, Kirkpatrick, Padgett, McMillan, Skrutskie, Stanford, Cohen, Walker,
  Mather, Leisawitz, Gautier, McLean, Benford, Lonsdale, Blain, Mendez, Irace,
  Duval, Liu, Royer, Heinrichsen, Howard, Shannon, Kendall, Walsh, Larsen,
  Cardon, Schick, Schwalm, Abid, Fabinsky, Naes, \& Tsai}]{Wright2010}
Wright, E.~L., Eisenhardt, P. R.~M., Mainzer, A.~K., {et~al.} 2010, \aj, 140,
  1868, \dodoi{10.1088/0004-6256/140/6/1868}

\bibitem[{Xu {et~al.}(2010)Xu, Domingue, Cheng, Lu, Huang, Gao, Mazzarella,
  Cutri, Sun, \& Surace}]{Xu2010}
Xu, C.~K., Domingue, D., Cheng, Y.-W., {et~al.} 2010, ApJ, 713, 330,
  \dodoi{10.1088/0004-637X/713/1/330}

\bibitem[{Yoon {et~al.}(2017)Yoon, Chung, Smith, \& Jaff{\'e}}]{Yoon2017}
Yoon, H., Chung, A., Smith, R., \& Jaff{\'e}, Y.~L. 2017, \apj, 838, 81,
  \dodoi{10.3847/1538-4357/aa6579}

\bibitem[{Yu {et~al.}(2024)Yu, Fang, Xu, Feng, Feng, Gao, Jiang, \&
  Lisenfeld}]{Yu2024a}
Yu, Q., Fang, T., Xu, C.~K., {et~al.} 2024, \apjs, 273, 2,
  \dodoi{10.3847/1538-4365/ad4547}

\bibitem[{Zhang {et~al.}(2024)Zhang, Zhu, Jiang, Cheng, Wang, Wang, Xu, Liu,
  Yu, Qian, Yu, Ai, Jing, Xu, Liu, Guan, Sun, Yang, Huang, Hao, \& {FAST
  Collaboration}}]{Zhang2024}
Zhang, C.-P., Zhu, M., Jiang, P., {et~al.} 2024, Sci. China Phys. Mech.
  Astron., 67, 219511, \dodoi{10.1007/s11433-023-2219-7}

\bibitem[{Zibetti {et~al.}(2009)Zibetti, Charlot, \& Rix}]{Zibetti2009}
Zibetti, S., Charlot, S., \& Rix, H.-W. 2009, \mnras, 400, 1181,
  \dodoi{10.1111/j.1365-2966.2009.15528.x}

\bibitem[{Zuo {et~al.}(2018)Zuo, Xu, Yun, Lisenfeld, Li, \& Cao}]{Zuo2018}
Zuo, P., Xu, C.~K., Yun, M.~S., {et~al.} 2018, ApJS, 237, 2,
  \dodoi{10.3847/1538-4365/aabd30}

\end{thebibliography}
\bibliographystyle{aasjournal}


\end{CJK*}
\end{document}